\documentclass[conference]{IEEEtran}
\usepackage{ifpdf}
\usepackage[nospace]{cite}
\usepackage{graphicx}
\usepackage{color}
\usepackage{bm}
\usepackage[cmex10]{amsmath}
\usepackage{amssymb}
\usepackage{array}
\usepackage{syntonly}
\usepackage{amsthm}
\usepackage{amsfonts}
\usepackage{epsfig}
\usepackage{latexsym}
\newcommand{\Define}{\stackrel{\triangle}{=}}

\DeclareMathOperator{\Tr}{Tr}

\usepackage{fixltx2e}
\usepackage{url}
\theoremstyle{definition} 
\theoremstyle{definition} 
\theoremstyle{definition} 

\begin{document}
\title{{\LARGE Sum Secrecy Rate in MISO Full-Duplex Wiretap Channel 
with Imperfect CSI} }
\author{\IEEEauthorblockN{Sanjay Vishwakarma}
\IEEEauthorblockA{Dept. of ECE\\
Indian Institute of Science\\
Bangalore 560012 \\
sanjay@ece.iisc.ernet.in}
\and
\IEEEauthorblockN{A. Chockalingam}
\IEEEauthorblockA{Dept. of ECE\\
Indian Institute of Science\\
Bangalore 560012 \\
Email: achockal@ece.iisc.ernet.in}}
\vspace{-30mm}
\author{{\large Sanjay Vishwakarma and A. Chockalingam} \\
sanjay@ece.iisc.ernet.in, achockal@ece.iisc.ernet.in \\
{\normalsize Department of ECE, Indian Institute of Science,
Bangalore 560012}
}
\maketitle
\begin{abstract}
In this paper, we consider the achievable sum secrecy rate in MISO 
(multiple-input-single-output) {\em full-duplex} wiretap channel 
in the presence of a passive eavesdropper and imperfect channel 
state information (CSI). We assume that the users participating in 
full-duplex communication have multiple transmit antennas, and that 
the users and the eavesdropper have single receive antenna each. The 
users have individual transmit power constraints. They also transmit 
jamming signals to improve the secrecy rates. We obtain the achievable 
perfect secrecy rate region by maximizing the worst case sum secrecy 
rate. We also obtain the corresponding transmit covariance matrices 
associated with the message signals and the jamming signals. Numerical 
results that show the impact of imperfect CSI on the achievable secrecy 
rate region are presented.
\end{abstract}
{\em keywords:}
{\em {\footnotesize
MISO, full-duplex, physical layer security, secrecy rate, 
semidefinite programming.
}} 
\IEEEpeerreviewmaketitle

\section{Introduction}
\label{sec1}
Transmitting messages with perfect secrecy using physical layer 
techniques was first studied in \cite{ic1} on a physically degraded 
discrete memoryless wiretap channel model. Later, this work was 
extended to more general broadcast channel in \cite{ic2} and Gaussian 
channel in \cite{ic3}, respectively. Wireless transmissions, being 
broadcast in nature, can be easily eavesdropped and hence require 
special attention to design modern secure wireless networks. Secrecy 
rate and capacity of point-to-point multi-antenna wiretap channels
have been reported in the literature by several authors, e.g., 
\cite{ic4, ic7, ic9, ic10}. In the above works, the 
transceiver operates in half-duplex mode, i.e., either it transmits 
or receives at any given time instant. On the other hand, full-duplex 
operation gives the advantage of simultaneous transmission and reception 
of messages \cite{fd_rice}. But loopback self-interference and imperfect 
channel state information (CSI) are limitations. Full-duplex communication 
without secrecy constraint has been investigated by many authors, e.g., 
\cite{ic11, ic12, ic13, ic14}. Full-duplex communication with secrecy 
constraint has been investigated in \cite{ic20, ic21, ic22}, where the 
achievable secrecy rate region of two-way (i.e., full-duplex) 
Gaussian and discrete memoryless 
wiretap channels have been characterized. In the above works, CSI in all 
the links are assumed to be perfect.

In this paper, we consider the achievable sum secrecy rate in MISO 
{\em full-duplex} wiretap channel in the presence of a passive 
eavesdropper and imperfect CSI. The users participating in full-duplex 
communication have multiple transmit antennas, and single receive 
antenna each. The eavesdropper is assumed to have single receive 
antenna. The norm of the CSI errors in all the links are assumed to 
be bounded in their respective absolute values. In addition to a 
message signal, each user transmits a jamming signal in order to 
improve the secrecy rates. The users operate under individual power 
constraints. For this scenario, we obtain the achievable perfect secrecy 
rate region by maximizing the worst case sum secrecy rate. We also obtain 
the corresponding transmit covariance matrices associated with the message 
signals and the jamming signals. Numerical results that illustrate the 
impact of imperfect CSI on the achievable secrecy rate region are presented.
We also minimize the total transmit power (sum of the transmit powers of 
users 1 and 2) with imperfect CSI subject to receive 
signal-to-interference-plus-noise ratio (SINR)
constraints at the users and eavesdropper, and individual transmit 
power constraints of the users.

The rest of the paper is organized as follows. The system model is
given in Sec. \ref{sec2}. Secrecy rate for perfect CSI is presented in
Sec. \ref{sec3}. Secrecy rate with imperfect CSI is studied in Sec. 
\ref{sec4}. Results and discussions are presented in Sec. \ref{sec6}. 
Conclusions are presented in Sec. \ref{sec7}. 

$\bf{Notations:}$ $\boldsymbol{A} \in 
\mathbb{C}^{N_{1} \times N_{2}}$ implies that $\boldsymbol{A}$ is a 
complex matrix of dimension $N_{1} \times N_{2}$. 
$\boldsymbol{A} \succeq \boldsymbol{0}$ and 
$\boldsymbol{A} \succ \boldsymbol{0}$ imply
that $\boldsymbol{A}$ is a positive semidefinite matrix and
positive definite matrix, respectively.
Identity matrix is denoted by $\boldsymbol{I}$.
$[.]^{\ast}$ denotes complex conjugate 
transpose operation.
$\mathbb{E}[.]$ denotes expectation operator. 
$\parallel\hspace{-1mm}.\hspace{-1mm}\parallel$ denotes 2-norm operator.
Trace of matrix $\boldsymbol{A} \in \mathbb{C}^{N \times N}$ is denoted 
by $\Tr(\boldsymbol{A})$.

\section{System Model}
\label{sec2}
\begin{figure}
\center
\includegraphics[totalheight=5.0cm,width=8.5cm]{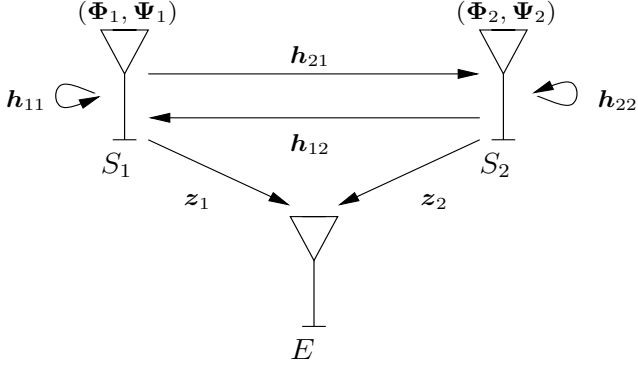}
\caption{System model for MISO full-duplex communication. 
$S_{1}$ has $M_{1}$ transmit antennas and single receive antenna.
$S_{2}$ has $M_{2}$ transmit antennas and single receive antenna.
$E$ has single receive antenna.}
\label{fig1}
\end{figure}

We consider full-duplex communication between two users $S_{1}$ and $S_{2}$
in the presence of an eavesdropper $E_{}$. $S_{1}$, $S_{2}$ are 
assumed to have $M_{1}$ and $M_{2}$ transmit antennas, respectively, and
single receive antenna each. $E$ is a passive eavesdropper
and it has single receive antenna. The complex channel gains on various 
links are as shown in Fig. \ref{fig1}, where
$\boldsymbol{h}_{11} \in \mathbb{C}^{1 \times M_{1}}$,
$\boldsymbol{h}_{12} \in \mathbb{C}^{1 \times M_{2}}$,
$\boldsymbol{h}_{21} \in \mathbb{C}^{1 \times M_{1}}$,
$\boldsymbol{h}_{22} \in \mathbb{C}^{1 \times M_{2}}$,
$\boldsymbol{z}_{1}  \in \mathbb{C}^{1 \times M_{1}}$, and
$\boldsymbol{z}_{2}  \in \mathbb{C}^{1 \times M_{2}}$.
$S_{1}$ and $S_{2}$ simultaneously transmit messages $W_{1}$ and $W_{2}$, 
respectively, in $n$ channel uses. $W_{1}$ and $W_{2}$ are independent and 
equiprobable over $\{1,2,\cdots,2^{nR_{1}}\}$ and $\{1,2,\cdots,2^{nR_{2}}\}$,
respectively. $R_{1}$ and $R_{2}$ are the information rates (bits per channel 
use) associated with $W_{1}$ and $W_{2}$, respectively, which need to be 
transmitted with perfect secrecy with respect to $E_{}$ \cite{ic22}. 
$S_{1}$ and $S_{2}$ map $W_{1}$ and $W_{2}$ to codewords 
$\{\boldsymbol{x}^{}_{1i}\}^{n}_{i = 1}$ $ \big (\boldsymbol{x}^{}_{1i} \in \mathbb{C}^{M_{1} \times 1}$, i.i.d. 
$ \sim \mathcal{CN}( \boldsymbol{0}, \boldsymbol{\Phi}_{1})$, 
$\boldsymbol{\Phi}_{1} = \mathbb{E}[\boldsymbol{x}^{}_{1i} \boldsymbol{x}^{\ast}_{1i}] \big )$ 
and $\{ \boldsymbol{x}^{}_{2i}\}^{n}_{i = 1}$ $ \big ( \boldsymbol{x}^{}_{2i} \in \mathbb{C}^{M_{2} \times 1}$, i.i.d. 
$ \sim \mathcal{CN}( \boldsymbol{0}, \boldsymbol{\Phi}_{2})$, 
$\boldsymbol{\Phi}_{2} = \mathbb{E}[\boldsymbol{x}^{}_{2i} \boldsymbol{x}^{\ast}_{2i}] \big )$, 
respectively, of length $n$. In order to degrade the eavesdropper channels 
and improve the secrecy rates, both $S_{1}$ and $S_{2}$ inject jamming signals
$\{ \boldsymbol{n}^{}_{1i}\}^{n}_{i = 1}$ $ \big( \boldsymbol{n}^{}_{1i} \in \mathbb{C}^{M_{1} \times 1}$, 
i.i.d. $ \sim \mathcal{CN}( \boldsymbol{0}, \boldsymbol{\Psi}_{1})$, 
$\boldsymbol{\Psi}_{1} = \mathbb{E}[\boldsymbol{n}^{}_{1i} \boldsymbol{n}^{\ast}_{1i}] \big )$ and
$\{\boldsymbol{n}^{}_{2i}\}^{n}_{i = 1}$ $ \big (\boldsymbol{n}^{}_{2i} \in \mathbb{C}^{M_{2} \times 1}$, 
i.i.d. $ \sim \mathcal{CN}( \boldsymbol{0}, \boldsymbol{\Psi}_{2})$, 
$\boldsymbol{\Psi}_{2} = \mathbb{E}[\boldsymbol{n}^{}_{2i} \boldsymbol{n}^{\ast}_{2i}] \big )$, 
respectively, of length $n$.
$S_{1}$ and $S_{2}$ transmit the symbols 
$\boldsymbol{x}^{}_{1i} + \boldsymbol{n}^{}_{1i}$ and 
$\boldsymbol{x}^{}_{2i} + \boldsymbol{n}^{}_{2i}$, respectively, during
the $i$th channel use, $1 \leq i \leq n$.
Hereafter, we will denote the 
symbols in $\{\boldsymbol{x}^{}_{1i}\}^{n}_{i = 1}$, $\{\boldsymbol{x}^{}_{2i}\}^{n}_{i = 1}$  
$\{ \boldsymbol{n}^{}_{1i}\}^{n}_{i = 1}$, and $\{ \boldsymbol{n}^{}_{2i}\}^{n}_{i = 1}$ by 
$\boldsymbol{x}^{}_{1}, \ \boldsymbol{x}^{}_{2}, \ \boldsymbol{n}^{}_{1} $, and $ \boldsymbol{n}^{}_{2}$, respectively.
We also assume that all the channel gains remain static over the codeword 
transmit duration. Let $P_{1}$ and $P_{2}$ be the transmit power budget for 
$S_{1}$ and $S_{2}$, respectively. This implies that 
\begin{eqnarray}
\Tr(\boldsymbol{\Phi}_{1} + \boldsymbol{\Psi}_{1}) \ \leq \ P_{1}, \ \ \Tr(\boldsymbol{\Phi}_{2} + \boldsymbol{\Psi}_{2}) \ \leq \ P_{2}. 
\label{eqn52}
\end{eqnarray}
Let $y_{1}$, $y_{2}$, and $y_{E}$ denote the received signals at $S_{1}$, 
$S_{2}$ and $E_{}$, respectively.
We have 
\begin{eqnarray}
y_{1}   &=& \boldsymbol{h}_{11}( \boldsymbol{x}^{}_{1}  + \boldsymbol{n}^{}_{1} ) +  \boldsymbol{h}_{12}(\boldsymbol{x}^{}_{2}   + \boldsymbol{n}^{}_{2})   + \eta_{1}, 
\label{eqn1} \\
y_{2}   &=& \boldsymbol{h}_{21}( \boldsymbol{x}^{}_{1}  + \boldsymbol{n}^{}_{1})  +  \boldsymbol{h}_{22}(\boldsymbol{x}^{}_{2} + \boldsymbol{n}^{}_{2})  + \eta_{2}, 
\label{eqn2} \\
y_{E}   &=& \boldsymbol{z}_{1}(  \boldsymbol{x}^{}_{1}  + \boldsymbol{n}^{}_{1})  +  \boldsymbol{z}_{2}(\boldsymbol{x}^{}_{2}  + \boldsymbol{n}^{}_{2})  + \eta_{E},        
\label{eqn3}
\end{eqnarray}
where $\eta_{1}$, $\eta_{2}$, and $\eta_{E}$ are i.i.d. 
$(\sim \mathcal{CN}(0, N_{0}))$ receiver noise terms.

\section{Sum secrecy rate - perfect CSI}
\label{sec3}
In this section, we assume perfect CSI in all the links. Since $S_{1}$ 
knows the transmitted symbol $(\boldsymbol{x}^{}_{1}+\boldsymbol{n}^{}_{1})$, 
in order to detect $\boldsymbol{x}^{}_{2}$, 
$S_{1}$ subtracts $\boldsymbol{h}_{11}( \boldsymbol{x}^{}_{1}+\boldsymbol{n}^{}_{1} )$ 
from the received signal $y_{1}$, i.e.,
\begin{eqnarray}
y^{'}_{1} \ &=& \ y_{1} - \boldsymbol{h}_{11}( \boldsymbol{x}^{}_{1}  + \boldsymbol{n}^{}_{1} ) \nonumber \\
            &=& \ \boldsymbol{h}_{12}(\boldsymbol{x}^{}_{2}   + \boldsymbol{n}^{}_{2}) + \eta_{1}. 
\label{eqn8} 
\end{eqnarray}
Similarly, since $S_{2}$ knows the transmitted symbol 
$( \boldsymbol{x}^{}_{2} + \boldsymbol{n}^{}_{2})$, 
to detect $\boldsymbol{x}^{}_{1}$, $S_{2}$ subtracts 
$\boldsymbol{h}_{22}( \boldsymbol{x}^{}_{2} + \boldsymbol{n}^{}_{2})$ 
from the received signal $y_{2}$, i.e., 
\begin{eqnarray}
y^{'}_{2} \ &=& \ y_{2} - \boldsymbol{h}_{22}( \boldsymbol{x}^{}_{2} + \boldsymbol{n}^{}_{2}) \nonumber \\
\ &=& \ \boldsymbol{h}_{21}( \boldsymbol{x}^{}_{1} + \boldsymbol{n}^{}_{1}) + \eta_{2}. 
\label{eqn9} 
\end{eqnarray}
Using (\ref{eqn8}) and (\ref{eqn9}), we get the following information rates 
for $\boldsymbol{x}^{}_{1}$ and $\boldsymbol{x}^{}_{2}$, respectively:
\begin{eqnarray}
R^{'}_{1} \Define I\big(\boldsymbol{x}^{}_{1}; \ y^{'}_{2}\big) = \log_{2} \Big( 1 + \frac{ \boldsymbol{h}_{21} \boldsymbol{\Phi}_{1} \boldsymbol{h}^{\ast}_{21} }{N_{0} + \boldsymbol{h}_{21} \boldsymbol{\Psi}_{1}  \boldsymbol{h}^{\ast}_{21} }\Big),
\label{eqn54} \\
R^{'}_{2} \Define I\big(\boldsymbol{x}^{}_{2}; \ y^{'}_{1}\big) = \log_{2} \Big( 1 + \frac{ \boldsymbol{h}_{12} \boldsymbol{\Phi}_{2} \boldsymbol{h}^{\ast}_{12} }{N_{0} + \boldsymbol{h}_{12} \boldsymbol{\Psi}_{2}  \boldsymbol{h}^{\ast}_{12} }\Big).
\label{eqn53} 
\end{eqnarray}
Using (\ref{eqn3}), we get the information leakage rate at $E$ as 
\begin{eqnarray}
R^{'}_{E} & \Define & I\big(\boldsymbol{x}^{}_{1}, \boldsymbol{x}^{}_{2}; \   y^{}_{E}\big) \nonumber \\ 
& = & \log_{2} \Big( 1 + \frac{ \boldsymbol{z}_{1} \boldsymbol{\Phi}_{1} \boldsymbol{z}^{\ast}_{1} + \boldsymbol{z}_{2} \boldsymbol{\Phi}_{2} \boldsymbol{z}^{\ast}_{2} }{ N_{0} + \boldsymbol{z}_{1} \boldsymbol{\Psi}_{1}  \boldsymbol{z}^{\ast}_{1} + \boldsymbol{z}_{2} \boldsymbol{\Psi}_{2} \boldsymbol{z}^{\ast}_{2} } \Big). 
\label{eqn55}
\end{eqnarray}
Using (\ref{eqn54}), (\ref{eqn53}), and (\ref{eqn55}), we get the 
information capacities
$C^{'}_{1}$, $C^{'}_{2}$, and $C^{'}_{E}$, respectively, as follows:
\begin{eqnarray}
C^{'}_{1} \  = \ \log_{2} \Big( 1 + \frac{ \parallel \hspace{-1mm} \boldsymbol{h}_{21} \hspace{-1mm} \parallel^{2} P_{1} }{N_{0} }\Big) \label{eqn71}, \\
C^{'}_{2} \  = \  \log_{2} \Big( 1 + \frac{ \parallel \hspace{-1mm} \boldsymbol{h}_{12} \hspace{-1mm} \parallel^{2} P_{2}}{N_{0}}\Big) \label{eqn72}, \\
C^{'}_{E} \  = \  \log_{2} \Big( 1 + \frac{ \parallel \hspace{-1mm} \boldsymbol{z}_{1} \hspace{-1mm} \parallel^{2} P_{1} + \parallel \hspace{-1mm} \boldsymbol{z}_{2} \hspace{-1mm} \parallel^{2} P_{2} }{ N_{0} } \Big).
\end{eqnarray}
A secrecy rate pair $(R_{1}, R_{2})$ which falls in the following region 
is achievable \cite{ic22}:
\begin{eqnarray}
0 \ \leq \ R_{1} \ \leq  \ R^{'}_{1}, \quad
0 \ \leq \ R_{2} \ \leq  \ R^{'}_{2}, \nonumber \\
0 \ \leq \ R_{1} + R_{2} \ \leq \ R^{'}_{1} + R^{'}_{2} - R^{'}_{E}, \nonumber \\
\boldsymbol{\Phi}_{1} \succeq \boldsymbol{0}, \quad \boldsymbol{\Psi}_{1} \succeq \boldsymbol{0}, \quad \Tr(\boldsymbol{\Phi}_{1} + \boldsymbol{\Psi}_{1}) \ \leq \ P_{1}, \nonumber \\
\boldsymbol{\Phi}_{2} \succeq \boldsymbol{0}, \quad \boldsymbol{\Psi}_{2} \succeq \boldsymbol{0}, \quad \Tr(\boldsymbol{\Phi}_{2} + \boldsymbol{\Psi}_{2}) \ \leq \ P_{2}.
\label{eqn73}
\end{eqnarray}
We intend to maximize the sum secrecy rate subject to the power 
constraint, i.e.,
\begin{eqnarray}
\max_{\boldsymbol{\Phi}_{1}, \ \boldsymbol{\Psi}_{1}, \atop{\boldsymbol{\Phi}_{2}, \ \boldsymbol{\Psi}_{2}}} \ R^{'}_{1} + R^{'}_{2} - R^{'}_{E} & & \label{eqn74} \\
 =   \max_{\boldsymbol{\Phi}_{1}, \ \boldsymbol{\Psi}_{1}, \atop{\boldsymbol{\Phi}_{2}, \ \boldsymbol{\Psi}_{2}}} \ \Big \{ \log_{2} \Big( 1 + \frac{ \boldsymbol{h}_{21} \boldsymbol{\Phi}_{1} \boldsymbol{h}^{\ast}_{21} }{N_{0} + \boldsymbol{h}_{21} \boldsymbol{\Psi}_{1}  \boldsymbol{h}^{\ast}_{21} }\Big) \nonumber \\ 
+ \log_{2} \Big( 1 + \frac{ \boldsymbol{h}_{12} \boldsymbol{\Phi}_{2} \boldsymbol{h}^{\ast}_{12} }{N_{0} + \boldsymbol{h}_{12} \boldsymbol{\Psi}_{2} \boldsymbol{h}^{\ast}_{12} }\Big) \nonumber \\ 
- \log_{2} \Big( 1 + \frac{ \boldsymbol{z}_{1} \boldsymbol{\Phi}_{1} \boldsymbol{z}^{\ast}_{1} + \boldsymbol{z}_{2} \boldsymbol{\Phi}_{2} \boldsymbol{z}^{\ast}_{2} }{ N_{0} + \boldsymbol{z}_{1} \boldsymbol{\Psi}_{1} \boldsymbol{z}^{\ast}_{1} + \boldsymbol{z}_{2} \boldsymbol{\Psi}_{2} \boldsymbol{z}^{\ast}_{2} } \Big) \Big \} \label{eqn75}
\\
\text{s.t.} \quad \boldsymbol{\Phi}_{1} \succeq \boldsymbol{0}, \quad \boldsymbol{\Psi}_{1} \succeq \boldsymbol{0}, \quad \Tr(\boldsymbol{\Phi}_{1} + \boldsymbol{\Psi}_{1}) \ \leq \ P_{1}, \nonumber \\
\boldsymbol{\Phi}_{2} \succeq \boldsymbol{0}, \quad \boldsymbol{\Psi}_{2} \succeq \boldsymbol{0}, \quad \Tr(\boldsymbol{\Phi}_{2} + \boldsymbol{\Psi}_{2}) \ \leq \ P_{2}.
\label{eqn76}
\end{eqnarray}
This is a non-convex optimization problem, and we solve it using 
two-dimensional search as follows.

$\textbf{Step1 :}$ Divide the intervals $[0, C^{'}_1]$ and $[0, C^{'}_2]$ 
in $K$ and 
$L$ small intervals, respectively, of size $\triangle_{1} = \frac{C^{'}_1}{K}$ 
and $\triangle_{2} = \frac{C^{'}_2}{L}$ where $K$ and $L$ are large integers.
Let $R^{'k}_{1} = k \triangle_{1}$ and $R^{'l}_{2} = l \triangle_{2}$, 
where $k = 0,1,2,\cdots,K$ and $l = 0,1,2,\cdots,L$.

$\textbf{Step2 :}$ For a given $(R^{'k}_{1}, \ R^{'l}_{2})$ pair, we 
minimize $R^{'}_{E}$ as follows:
\begin{eqnarray}
R^{''kl}_{E} \Define \min_{\boldsymbol{\Phi}_{1}, \ \boldsymbol{\psi}_{1}, \atop{\boldsymbol{\Phi}_{2}, \ \boldsymbol{\psi}_{2}}} \ \log_{2} \Big( 1 + \frac{ \boldsymbol{z}_{1} \boldsymbol{\Phi}_{1}  \boldsymbol{z}^{\ast}_{1} + \boldsymbol{z}_{2} \boldsymbol{\Phi}_{2} \boldsymbol{z}^{\ast}_{2} }{ N_{0} + \boldsymbol{z}_{1} \boldsymbol{\Psi}_{1} \boldsymbol{z}^{\ast}_{1} + \boldsymbol{z}_{2} \boldsymbol{\Psi}_{2} \boldsymbol{z}^{\ast}_{2} } \Big) \label{eqn77} 
\\
\text{s.t.} \quad R^{''k}_{1} \ \Define \ \log_{2} \Big( 1 + \frac{ \boldsymbol{h}_{21} \boldsymbol{\Phi}_{1}  \boldsymbol{h}^{\ast}_{21} }{N_{0} + \boldsymbol{h}_{21} \boldsymbol{\Psi}_{1}  \boldsymbol{h}^{\ast}_{21} }\Big) \ \geq \ R^{'k}_{1}, \nonumber \\
R^{''l}_{2} \ \Define \  \log_{2} \Big( 1 + \frac{ \boldsymbol{h}_{12} \boldsymbol{\Phi}_{2}  \boldsymbol{h}^{\ast}_{12} }{N_{0} + \boldsymbol{h}_{12} \boldsymbol{\Psi}_{2}  \boldsymbol{h}^{\ast}_{12} }\Big) \ \geq \ R^{'l}_{2}, \nonumber \\
\boldsymbol{\Phi}_{1} \succeq \boldsymbol{0}, \quad \boldsymbol{\Psi}_{1} \succeq \boldsymbol{0}, \quad \Tr(\boldsymbol{\Phi}_{1} + \boldsymbol{\Psi}_{1}) \ \leq \ P_{1}, \nonumber \\
\boldsymbol{\Phi}_{2} \succeq \boldsymbol{0}, \quad \boldsymbol{\Psi}_{2} \succeq \boldsymbol{0}, \quad \Tr(\boldsymbol{\Phi}_{2} + \boldsymbol{\Psi}_{2}) \ \leq \ P_{2}.
\end{eqnarray}
The maximum sum secrecy rate is given by 
$\max_{k = 0,1,2,\cdots,K, \atop{l = 0,1,2,\cdots,L}} \ (R^{''k}_{1} + R^{''l}_{2} - R^{''kl}_{E})$.
We solve the optimization problem (\ref{eqn77}) as follows.
Dropping the logarithm in the objective function in (\ref{eqn77}), we 
rewrite the optimization problem (\ref{eqn77}) in the following equivalent 
form:
\begin{eqnarray}
\min_{t, \ \boldsymbol{\Phi}_{1}, \ \boldsymbol{\Psi}_{1}, \ \boldsymbol{\Phi}_{2}, \ \boldsymbol{\Psi}_{2}} \ \ t 
\label{eqn79} 
\end{eqnarray}
s.t.
\vspace{-1mm}
{\small
\begin{eqnarray}
%\mbox{s.t.} \ \ 
\big( \boldsymbol{z}_{1} \boldsymbol{\Phi}_{1}  \boldsymbol{z}^{\ast}_{1} + \boldsymbol{z}_{2} \boldsymbol{\Phi}_{2} \boldsymbol{z}^{\ast}_{2} \big) - t \big( N_{0} + \boldsymbol{z}_{1} \boldsymbol{\Psi}_{1}  \boldsymbol{z}^{\ast}_{1} + \boldsymbol{z}_{2} \boldsymbol{\Psi}_{2} \boldsymbol{z}^{\ast}_{2} \big) \ \leq \ 0, \nonumber \\
\big( 2^{R^{'k}_{1}} - 1 \big) \big( N_{0} + \boldsymbol{h}_{21} \boldsymbol{\Psi}_{1}  \boldsymbol{h}^{\ast}_{21} \big) - \big( \boldsymbol{h}_{21} \boldsymbol{\Phi}_{1}  \boldsymbol{h}^{\ast}_{21} \big) \ \leq \ 0, \nonumber \\
\big( 2^{R^{'l}_{2}} - 1 \big) \big( N_{0} + \boldsymbol{h}_{12} \boldsymbol{\Psi}_{2}  \boldsymbol{h}^{\ast}_{12} \big) - \big( \boldsymbol{h}_{12} \boldsymbol{\Phi}_{2}  \boldsymbol{h}^{\ast}_{12} \big) \ \leq \ 0, \nonumber \\
\boldsymbol{\Phi}_{1} \succeq \boldsymbol{0}, \quad \boldsymbol{\Psi}_{1} \succeq \boldsymbol{0}, \quad \Tr(\boldsymbol{\Phi}_{1} + \boldsymbol{\Psi}_{1}) \ \leq \ P_{1}, \nonumber \\
\boldsymbol{\Phi}_{2} \succeq \boldsymbol{0}, \quad \boldsymbol{\Psi}_{2} \succeq \boldsymbol{0}, \quad \Tr(\boldsymbol{\Phi}_{2} + \boldsymbol{\Psi}_{2}) \ \leq \ P_{2}. 
\label{eqn80}
\end{eqnarray}
}

\vspace{-4mm}
\hspace{-4.5mm}
Using the KKT conditions of the above optimization problem, we analyze 
the ranks of the optimum solutions 
$\boldsymbol{\Phi}_{1}$, $\boldsymbol{\Psi}_{1}$,
$\boldsymbol{\Phi}_{2}$, $\boldsymbol{\Psi}_{2}$ in the Appendix.
Further, for a given $t$, the above problem is formulated as the 
following semidefinite feasibility problem \cite{ic16}: 
\begin{eqnarray}
\text{find} \quad \boldsymbol{\Phi}_{1}, \ \boldsymbol{\Psi}_{1}, \ \boldsymbol{\Phi}_{2}, \ \boldsymbol{\Psi}_{2} \label{eqn81}
\end{eqnarray}
subject to the constraints in (\ref{eqn80}). 
The minimum value of $t$, denoted by $t^{kl}_{min}$, can be obtained 
using bisection method \cite{ic16} as follows. Let $t^{kl}_{min}$ lie 
in the interval $[t_{lowerlimit}, \ t_{upperlimit}]$. The value of 
$t_{lowerlimit}$ can be taken as 0 (corresponding to the minimum 
information rate of 0) and $t_{upperlimit}$ can be taken as 
$(2^{C^{'}_{E}} - 1)$, which corresponds to the 
information capacity of the eavesdropper link.
Check the feasibility of 
(\ref{eqn81}) at $t^{kl}_{min} = (t^{}_{lowerlimit} + t^{}_{upperlimit})/2$. 
If feasible, then $t^{}_{upperlimit} = t^{kl}_{min}$, else 
$ \ t^{}_{lowerlimit} = t^{kl}_{min}$. Repeat this until 
$t^{}_{upperlimit} - t^{}_{lowerlimit} \leq \zeta$, where $\zeta$ is 
a small positive number. Using $t^{kl}_{min}$ in (\ref{eqn77}), 
$R^{''kl}_{E}$ is given by
\begin{eqnarray}
R^{''kl}_{E} \ = \ \log_2 (1 + t^{kl}_{min}). \label{eqn82}
\end{eqnarray}

\section{Sum secrecy rate - imperfect CSI}
\label{sec4}
In this section, we assume that the available CSI in all the links are 
imperfect \cite{ia50, ia51, ia52}, i.e.,
\begin{eqnarray}
\boldsymbol{h}_{11} = \boldsymbol{h}^{0}_{11} + \boldsymbol{e}_{11}, \quad \boldsymbol{h}_{12} = \boldsymbol{h}^{0}_{12} + \boldsymbol{e}_{12},  \quad
\boldsymbol{h}_{21} = \boldsymbol{h}^{0}_{21}+\boldsymbol{e}_{21}, \nonumber \\ 
\boldsymbol{h}_{22} = \boldsymbol{h}^{0}_{22} + \boldsymbol{e}_{22},  \quad
\boldsymbol{z}_{1}  = \boldsymbol{z}^{0}_{1}  + \boldsymbol{e}_{1},  \quad \boldsymbol{z}_{2}  = \boldsymbol{z}^{0}_{2}  + \boldsymbol{e}_{2},   
\nonumber 
\end{eqnarray}
where $\boldsymbol{h}^{0}_{11}$, $\boldsymbol{h}^{0}_{12}$, 
$\boldsymbol{h}^{0}_{21}$, $\boldsymbol{h}^{0}_{22}$, 
$\boldsymbol{z}^{0}_{1}$, and  $\boldsymbol{z}^{0}_{2}$ are the 
estimates of $\boldsymbol{h}^{}_{11}$, $\boldsymbol{h}^{}_{12}$, 
$\boldsymbol{h}^{}_{21}$, $\boldsymbol{h}^{}_{22}$, 
$\boldsymbol{z}^{}_{1}$, and  $\boldsymbol{z}^{}_{2}$, respectively, 
and $\boldsymbol{e}^{}_{11}$, $\boldsymbol{e}^{}_{12}$, 
$\boldsymbol{e}^{}_{21}$, $\boldsymbol{e}^{}_{22}$, 
$\boldsymbol{e}^{}_{1}$, and  $\boldsymbol{e}^{}_{2}$
are the corresponding errors. We assume that the norm of the errors 
are bounded in their respective absolute values as: 
\begin{eqnarray}
{\parallel \hspace{-1mm} \boldsymbol{e}_{11} \hspace{-1mm} \parallel}^{} \leq {\epsilon}^{}_{11}, \quad {\parallel \hspace{-1mm} \boldsymbol{e}_{12} \hspace{-1mm} \parallel}^{} \leq {\epsilon}^{}_{12}, \quad%\label{eqn86} \\
{\parallel \hspace{-1mm} \boldsymbol{e}_{21} \hspace{-1mm} \parallel}^{} \leq {\epsilon}^{}_{21}, \nonumber \\
{\parallel \hspace{-1mm} \boldsymbol{e}_{22} \hspace{-1mm} \parallel}^{} \leq {\epsilon}^{}_{22}, \quad %\label{eqn87} \\
{\parallel \hspace{-1mm} \boldsymbol{e}_{1}  \hspace{-1mm} \parallel}^{} \leq {\epsilon}^{}_{1},  \quad {\parallel \hspace{-1mm} \boldsymbol{e}_{2}   \hspace{-1mm} \parallel}^{} \leq {\epsilon}^{}_{2}. \nonumber % \label{eqn88} 
\end{eqnarray}
We make the following assumptions with respect to the availability 
of the CSI at $S_{1}$, $S_{2}$, and $E$:

(a.) We assume that only the estimates $\boldsymbol{h}^{0}_{11}$, 
$\boldsymbol{h}^{0}_{21}$, $\boldsymbol{z}^{0}_{1}$, and
$\boldsymbol{z}^{0}_{2}$ are available at $S_{1}$ while 
$\boldsymbol{h}^{}_{12}$ is perfectly known at $S_{1}$ (coherent 
detection). Similarly, only the estimates $\boldsymbol{h}^{0}_{22}$, 
$\boldsymbol{h}^{0}_{12}$, $\boldsymbol{z}^{0}_{1}$, and
$\boldsymbol{z}^{0}_{2}$ are available at $S_{2}$ while 
$\boldsymbol{h}^{}_{21}$ is perfectly known at $S_{2}$ (coherent 
detection). We assume that $E$ has perfect knowledge of 
$\boldsymbol{z}^{}_{1}$, and $\boldsymbol{z}^{}_{2}$ (coherent 
detection). With the above error model, we rewrite (\ref{eqn8}), 
(\ref{eqn9}), and (\ref{eqn3}) as follows:
\begin{eqnarray}
y^{'}_{1} \ &=& \ y_{1} - \boldsymbol{h}^{0}_{11}(\boldsymbol{x}^{}_{1} + \boldsymbol{n}^{}_{1}) \nonumber \\ 
&=& \boldsymbol{e}^{}_{11}(\boldsymbol{x}^{}_{1} + \boldsymbol{n}^{}_{1}) + \boldsymbol{h}_{12}(\boldsymbol{x}^{}_{2} + \boldsymbol{n}^{}_{2}) + \eta_{1}, \label{eqn89} \\
\vspace{2mm}
y^{'}_{2} \ &=& \ y_{2} - \boldsymbol{h}^{0}_{22}(\boldsymbol{x}^{}_{2} + \boldsymbol{n}^{}_{2}) \nonumber \\ 
&=& \boldsymbol{h}_{21}(\boldsymbol{x}^{}_{1} + \boldsymbol{n}^{}_{1}) + \boldsymbol{e}^{}_{22}(\boldsymbol{x}^{}_{2} + \boldsymbol{n}^{}_{2}) +  \eta_{2}, \label{eqn90} \\
\vspace{2mm}
y_{E}     \ &=& \ \boldsymbol{z}_{1}(\boldsymbol{x}^{}_{1} + \boldsymbol{n}^{}_{1}) +  \boldsymbol{z}_{2}(\boldsymbol{x}^{}_{2} + \boldsymbol{n}^{}_{2})  + \eta_{E}. \label{eqn91} 
\end{eqnarray}

(b.) We assume that while detecting $\boldsymbol{x}^{}_{2}$, $S_{1}$ 
treats the residual term 
$\boldsymbol{e}^{}_{11}(\boldsymbol{x}^{}_{1} + \boldsymbol{n}^{}_{1})$ 
which appears in (\ref{eqn89}) as self-noise.
Similarly, while detecting $\boldsymbol{x}^{}_{1}$, $S_{2}$ treats the 
residual term 
$\boldsymbol{e}^{}_{22}(\boldsymbol{x}^{}_{2} + \boldsymbol{n}^{}_{2})$ 
which appears in (\ref{eqn90}) as self-noise.

Further, in order to compute $R^{'k}_{1}$, $R^{'l}_{2}$, and $R^{''kl}_{E}$, 
respectively, as described in $\textbf{Step1}$ and $\textbf{Step2}$ in 
Section \ref{sec3}, we get the worst case capacities $C^{'}_{1}$, 
$C^{'}_{2}$ for $S_{1}$, $S_{2}$ links, and best case capacity 
$C^{'}_{E}$ for the eavesdropper link with imperfect CSI as follows:

\vspace{-4mm}
{\small
\begin{eqnarray}
C^{'}_{1} 
= \log_{2} \Big( 1 + \frac{{\lvert \parallel \hspace{-1mm} \boldsymbol{h}^{0}_{21} \hspace{-1mm} \parallel - \epsilon_{21}\rvert}^{2}P_{1}}{N_{0}}\Big) \ \ \text{if} \ \ \big(\parallel \hspace{-1mm} \boldsymbol{h}^{0}_{21} \hspace{-1mm} \parallel \ > \ \epsilon_{21} \big), \nonumber \\ 0 \ \ \text{else}. \label{eqn100}  \\
C^{'}_{2} 
= \log_{2} \Big( 1 + \frac{{\lvert \parallel \hspace{-1mm} \boldsymbol{h}^{0}_{12} \hspace{-1mm} \parallel - \epsilon_{12}\rvert}^{2}P_{2}}{N_{0}}\Big) \ \ \text{if} \ \ \big(\parallel \hspace{-1mm} \boldsymbol{h}^{0}_{12} \hspace{-1mm} \parallel \ > \ \epsilon_{12} \big), \nonumber \\ 0 \ \ \text{else}.\label{eqn101}  \\
C^{'}_{E} 
= \log_{2} \Big( 1 + \frac{ {\lvert \parallel \hspace{-1mm} \boldsymbol{z}^{0}_{1} \hspace{-1mm} \parallel +  \epsilon_{1}  \rvert}^{2}P_{1} + {\lvert \parallel \hspace{-1mm} \boldsymbol{z}^{0}_{2} \hspace{-1mm} \parallel  +  \epsilon_{2} \rvert}^{2}P_{2} }{ N_{0} } \Big). \label{102}
\end{eqnarray}
}

\vspace{-4mm}
\hspace{-4.5mm}
Using (\ref{eqn89}), (\ref{eqn90}), and (\ref{eqn91}), we write the 
optimization problem (\ref{eqn77}) with imperfect CSI as follows:

\vspace{-1mm}
{\scriptsize
\begin{eqnarray}
R^{''kl}_{E}  \ \Define \ \min_{\boldsymbol{\Phi}_{1}, \ \boldsymbol{\Psi}_{1}, \ \boldsymbol{\Phi}_{2}, \ \boldsymbol{\Psi}_{2}} \quad \max_{\boldsymbol{e}_{1}, \ \boldsymbol{e}_{2}} \ \log_{2} \nonumber \\ 
\bigg( 1 + \frac{ (\boldsymbol{z}^{0}_{1} + \boldsymbol{e}_{1}) \boldsymbol{\Phi}_{1}  (\boldsymbol{z}^{0}_{1} + \boldsymbol{e}_{1} )^{\ast} + (\boldsymbol{z}^{0}_{2} + \boldsymbol{e}_{2} ) \boldsymbol{\Phi}_{2} (\boldsymbol{z}^{0}_{2} + \boldsymbol{e}_{2} )^{\ast} }{ N_{0} + (\boldsymbol{z}^{0}_{1} + \boldsymbol{e}_{1}) \boldsymbol{\Psi}_{1} (\boldsymbol{z}^{0}_{1} + \boldsymbol{e}_{1})^{\ast} + (\boldsymbol{z}^{0}_{2} + \boldsymbol{e}_{2}) \boldsymbol{\Psi}_{2} (\boldsymbol{z}^{0}_{2} + \boldsymbol{e}_{2})^{\ast} } \bigg) \label{eqn92}  
\end{eqnarray}
}

\vspace{-6mm}
{\scriptsize
\begin{eqnarray}
\text{s.t.} \quad \quad R^{''k}_{1} \ \Define \ \min_{\boldsymbol{e}_{21}, \ \boldsymbol{e}_{22}} \ \log_{2} \nonumber \\ 
\bigg( 1 + \frac{ (\boldsymbol{h}^{0}_{21} + \boldsymbol{e}_{21}) \boldsymbol{\Phi}_{1}  (\boldsymbol{h}^{0}_{21} + \boldsymbol{e}_{21})^{\ast} }{N_{0} + \boldsymbol{e}_{22}(\boldsymbol{\Phi}_{2} + \boldsymbol{\Psi}_{2}) \boldsymbol{e}^{\ast}_{22} + (\boldsymbol{h}^{0}_{21} + \boldsymbol{e}_{21}) \boldsymbol{\Psi}_{1}  (\boldsymbol{h}^{0}_{21} + \boldsymbol{e}_{21})^{\ast} }\bigg) \nonumber \\
\geq \ R^{'k}_{1}, \label{eqn93} \\
R^{''l}_{2} \ \Define \ \min_{\boldsymbol{e}_{11}, \ \boldsymbol{e}_{12}} \ \log_{2} \nonumber \\ 
\bigg( 1 + \frac{ (\boldsymbol{h}^{0}_{12} + \boldsymbol{e}_{12}) \boldsymbol{\Phi}_{2}  (\boldsymbol{h}^{0}_{12} + \boldsymbol{e}_{12})^{\ast} }{N_{0} + \boldsymbol{e}_{11}(\boldsymbol{\Phi}_{1} + \boldsymbol{\Psi}_{1}) \boldsymbol{e}^{\ast}_{11} + (\boldsymbol{h}^{0}_{12} + \boldsymbol{e}_{12}) \boldsymbol{\Psi}_{2}  (\boldsymbol{h}^{0}_{12} + \boldsymbol{e}_{12})^{\ast} }\bigg) \nonumber \\
\geq \ R^{'l}_{2}, \label{eqn94} \\
{\parallel \hspace{-1mm} \boldsymbol{e}_{11} \hspace{-1mm} \parallel}^{2} \leq {\epsilon}^{2}_{11}, \quad 
{\parallel \hspace{-1mm} \boldsymbol{e}_{12} \hspace{-1mm} \parallel}^{2} \leq {\epsilon}^{2}_{12}, \quad
{\parallel \hspace{-1mm} \boldsymbol{e}_{21} \hspace{-1mm} \parallel}^{2} \leq {\epsilon}^{2}_{21}, \nonumber \\
{\parallel \hspace{-1mm} \boldsymbol{e}_{22} \hspace{-1mm} \parallel}^{2} \leq {\epsilon}^{2}_{22}, \quad 
{\parallel \hspace{-1mm} \boldsymbol{e}_{1} \hspace{-1mm} \parallel}^{2} \leq {\epsilon}^{2}_{1}, \quad 
{\parallel \hspace{-1mm} \boldsymbol{e}_{2} \hspace{-1mm} \parallel}^{2} \leq {\epsilon}^{2}_{2}, \label{eqn95} \\
\boldsymbol{\Phi}_{1} \succeq \boldsymbol{0}, \quad \boldsymbol{\Psi}_{1} \succeq \boldsymbol{0}, \quad \Tr(\boldsymbol{\Phi}_{1} + \boldsymbol{\Psi}_{1}) \ \leq \ P_{1}, \nonumber \\
\boldsymbol{\Phi}_{2} \succeq \boldsymbol{0}, \quad \boldsymbol{\Psi}_{2} \succeq \boldsymbol{0}, \quad \Tr(\boldsymbol{\Phi}_{2} + \boldsymbol{\Psi}_{2}) \ \leq \ P_{2}.
\label{eqn96}
\end{eqnarray}
}

\vspace{-4mm}
\hspace{-5mm}
In the constraints (\ref{eqn93}) and (\ref{eqn94}), additional noise 
appear due the terms 
$\boldsymbol{e}^{}_{22}(\boldsymbol{x}^{}_{2} + \boldsymbol{n}^{}_{2})$ 
and $\boldsymbol{e}^{}_{11}(\boldsymbol{x}^{}_{1}+\boldsymbol{n}^{}_{1})$, 
respectively, which have been treated as self noise. 

We solve the optimization problem (\ref{eqn92}) as follows.
Dropping the logarithm in the objective function in (\ref{eqn92}), 
we write the optimization problem (\ref{eqn92}) in the following 
equivalent form:

\vspace{-8mm}
{\small
\begin{eqnarray}
\min_{\boldsymbol{\Phi}_{1}, \ \boldsymbol{\Psi}_{1}, \ \boldsymbol{\Phi}_{2}, \ \boldsymbol{\Psi}_{2}} \quad \max_{\boldsymbol{e}_{1}, \ \boldsymbol{e}_{2}} \nonumber \\ 
\bigg(\frac{ (\boldsymbol{z}^{0}_{1} + \boldsymbol{e}_{1}) \boldsymbol{\Phi}_{1}  (\boldsymbol{z}^{0}_{1} + \boldsymbol{e}_{1} )^{\ast} + (\boldsymbol{z}^{0}_{2} + \boldsymbol{e}_{2} ) \boldsymbol{\Phi}_{2} (\boldsymbol{z}^{0}_{2} + \boldsymbol{e}_{2} )^{\ast} }{ N_{0} + (\boldsymbol{z}^{0}_{1} + \boldsymbol{e}_{1}) \boldsymbol{\Psi}_{1} (\boldsymbol{z}^{0}_{1} + \boldsymbol{e}_{1})^{\ast} + (\boldsymbol{z}^{0}_{2} + \boldsymbol{e}_{2}) \boldsymbol{\Psi}_{2} (\boldsymbol{z}^{0}_{2} + \boldsymbol{e}_{2})^{\ast} } \bigg)
\label{eqn102}  \\
\vspace{2mm}
\text{s.t.} \quad \quad \min_{\boldsymbol{e}_{21}, \ \boldsymbol{e}_{22}} \nonumber \\ 
\bigg(\frac{ (\boldsymbol{h}^{0}_{21} + \boldsymbol{e}_{21}) \boldsymbol{\Phi}_{1}  (\boldsymbol{h}^{0}_{21} + \boldsymbol{e}_{21})^{\ast} }{N_{0} + \boldsymbol{e}_{22}(\boldsymbol{\Phi}_{2} + \boldsymbol{\Psi}_{2}) \boldsymbol{e}^{\ast}_{22} + (\boldsymbol{h}^{0}_{21} + \boldsymbol{e}_{21}) \boldsymbol{\Psi}_{1}  (\boldsymbol{h}^{0}_{21} + \boldsymbol{e}_{21})^{\ast} }\bigg) \nonumber \\ 
\geq \ (2^{R^{'k}_{1}} -1),  \label{eqn124} \\
\min_{\boldsymbol{e}_{11}, \ \boldsymbol{e}_{12}} \nonumber \\ 
\bigg(\frac{ (\boldsymbol{h}^{0}_{12} + \boldsymbol{e}_{12}) \boldsymbol{\Phi}_{2}  (\boldsymbol{h}^{0}_{12} + \boldsymbol{e}_{12})^{\ast} }{N_{0} + \boldsymbol{e}_{11}(\boldsymbol{\Phi}_{1} + \boldsymbol{\Psi}_{1}) \boldsymbol{e}^{\ast}_{11} + (\boldsymbol{h}^{0}_{12} + \boldsymbol{e}_{12}) \boldsymbol{\Psi}_{2}  (\boldsymbol{h}^{0}_{12} + \boldsymbol{e}_{12})^{\ast} }\bigg) \nonumber \\ 
\geq \ (2^{R^{'l}_{2}} -1),  \label{eqn125} \\
{\parallel \hspace{-1mm} \boldsymbol{e}_{11} \hspace{-1mm} \parallel}^{2} \leq {\epsilon}^{2}_{11}, \quad 
{\parallel \hspace{-1mm} \boldsymbol{e}_{12} \hspace{-1mm} \parallel}^{2} \leq {\epsilon}^{2}_{12}, \quad
{\parallel \hspace{-1mm} \boldsymbol{e}_{21} \hspace{-1mm} \parallel}^{2} \leq {\epsilon}^{2}_{21}, \nonumber \\
{\parallel \hspace{-1mm} \boldsymbol{e}_{22} \hspace{-1mm} \parallel}^{2} \leq {\epsilon}^{2}_{22}, \quad 
{\parallel \hspace{-1mm} \boldsymbol{e}_{1} \hspace{-1mm} \parallel}^{2} \leq {\epsilon}^{2}_{1}, \quad 
{\parallel \hspace{-1mm} \boldsymbol{e}_{2} \hspace{-1mm} \parallel}^{2} \leq {\epsilon}^{2}_{2},  \nonumber \\
\boldsymbol{\Phi}_{1} \succeq \boldsymbol{0}, \quad \boldsymbol{\Psi}_{1} \succeq \boldsymbol{0}, \quad \Tr(\boldsymbol{\Phi}_{1} + \boldsymbol{\Psi}_{1}) \ \leq \ P_{1}, \nonumber \\
\boldsymbol{\Phi}_{2} \succeq \boldsymbol{0}, \quad \boldsymbol{\Psi}_{2} \succeq \boldsymbol{0}, \quad \Tr(\boldsymbol{\Phi}_{2} + \boldsymbol{\Psi}_{2}) \ \leq \ P_{2}.
\label{eqn103}
\end{eqnarray}
}

\vspace{-5mm}
\hspace{-5mm}
Solving the above optimization problem is hard due to the presence of 
$\boldsymbol{e}_{1}$ and $\boldsymbol{e}_{2}$ in both the numerator and 
denominator of the objective function in (\ref{eqn102}). Similarly, 
$\boldsymbol{e}_{21}$ and $\boldsymbol{e}_{12}$ appear in both the 
numerator and denominator of the constraints in (\ref{eqn124}) and 
(\ref{eqn125}), respectively. By independently constraining the various 
quadratic terms appearing in the objective function in (\ref{eqn102})
and the constraints (\ref{eqn124}) and (\ref{eqn125}), we get the 
following upper bound for the above optimization problem:
\vspace{-4mm}
\begin{eqnarray}
\min_{\boldsymbol{\Phi}_{1}, \ \boldsymbol{\Psi}_{1}, \ \boldsymbol{\Phi}_{2}, \ \boldsymbol{\Psi}_{2} } \ \ \min_{t_{1}, \ t_{2}, \cdots,t_{10}} \ \Big( \frac{t_{1} + t_{2}}{N_{0} + t_{3} + t_{4}} \Big) \label{eqn104} 
\\
\text{s.t.} \quad t_{3} \ \geq \ 0, \ t_{4} \ \geq \ 0, \ t_{5} \ \geq \ 0, \ t_{8} \ \geq \ 0, \label{eqn110} \\
\forall \boldsymbol{e}_{1} \quad \text{s.t.} \quad {\parallel \hspace{-1mm} \boldsymbol{e}_{1} \hspace{-1mm}\parallel}^{2} \leq {\epsilon}^{2}_{1} \ \Longrightarrow \ \nonumber \\  
(\boldsymbol{z}^{0}_{1} + \boldsymbol{e}_{1}) \boldsymbol{\Phi}_{1} (\boldsymbol{z}^{0}_{1} + \boldsymbol{e}_{1})^{\ast} - t_{1} \ \leq \ 0, \label{eqn111} \\
\forall \boldsymbol{e}_{1} \quad \text{s.t.} \quad {\parallel \hspace{-1mm} \boldsymbol{e}_{1} \hspace{-1mm}\parallel}^{2} \leq {\epsilon}^{2}_{1} \ \Longrightarrow \ \nonumber \\ 
-(\boldsymbol{z}^{0}_{1} + \boldsymbol{e}_{1}) \boldsymbol{\Psi}_{1} (\boldsymbol{z}^{0}_{1} + \boldsymbol{e}_{1})^{\ast} + t_{3} \ \leq \ 0,  \label{eqn112} \\
\forall \boldsymbol{e}_{2} \quad \text{s.t.} \quad {\parallel \hspace{-1mm} \boldsymbol{e}_{2} \hspace{-1mm}\parallel}^{2} \leq {\epsilon}^{2}_{2} \ \Longrightarrow \ \nonumber \\  
(\boldsymbol{z}^{0}_{2} + \boldsymbol{e}_{2}) \boldsymbol{\Phi}_{2} (\boldsymbol{z}^{0}_{2} + \boldsymbol{e}_{2})^{\ast} - t_{2} \ \leq \ 0, \label{eqn113} \\
\forall \boldsymbol{e}_{2} \quad \text{s.t.} \quad {\parallel \hspace{-1mm} \boldsymbol{e}_{2} \hspace{-1mm}\parallel}^{2} \leq {\epsilon}^{2}_{2} \ \Longrightarrow \ \nonumber \\ 
-(\boldsymbol{z}^{0}_{2} + \boldsymbol{e}_{2}) \boldsymbol{\Psi}_{2} (\boldsymbol{z}^{0}_{2} + \boldsymbol{e}_{2})^{\ast} + t_{4} \ \leq \ 0, \label{eqn114} \\
\Big( \frac{t_{5}}{N_{0} + t_6 + t_{7}}\Big) \ \geq \ (2^{R^{'k}_{1}} -1),  \label{eqn115} \\
\forall \boldsymbol{e}_{21} \quad \text{s.t.} \quad {\parallel \hspace{-1mm} \boldsymbol{e}_{21} \hspace{-1mm} \parallel}^{2} \leq {\epsilon}^{2}_{21} \ \Longrightarrow \ \nonumber \\  
-(\boldsymbol{h}^{0}_{21} + \boldsymbol{e}_{21}) \boldsymbol{\Phi}_{1} (\boldsymbol{h}^{0}_{21} + \boldsymbol{e}_{21})^{\ast} + t_{5} \ \leq \ 0 ,  \label{eqn116} \\
\forall \boldsymbol{e}_{21} \quad \text{s.t.} \quad {\parallel \hspace{-1mm} \boldsymbol{e}_{21} \hspace{-1mm} \parallel}^{2} \leq {\epsilon}^{2}_{21} \ \Longrightarrow \ \nonumber \\ 
(\boldsymbol{h}^{0}_{21} + \boldsymbol{e}_{21}) \boldsymbol{\Psi}_{1} (\boldsymbol{h}^{0}_{21} + \boldsymbol{e}_{21})^{\ast} - t_{7} \ \leq \ 0, \label{eqn117} \\
\forall \boldsymbol{e}_{22} \quad \text{s.t.} \quad {\parallel \hspace{-1mm} \boldsymbol{e}_{22} \hspace{-1mm} \parallel}^{2} \leq {\epsilon}^{2}_{22} \ \Longrightarrow \ \nonumber \\ 
\boldsymbol{e}_{22} (\boldsymbol{\Phi}_{2} + \boldsymbol{\Psi}_{2}) \boldsymbol{e}^{\ast}_{22} - t_{6} \ \leq \ 0, \label{eqn126} \\
\Big( \frac{t_{8}}{N_{0} + t_9 + t_{10}}\Big) \ \geq \ (2^{R^{'l}_{2}} -1),  \label{eqn118} \\
\forall \boldsymbol{e}_{12} \quad \text{s.t.} \quad {\parallel \hspace{-1mm} \boldsymbol{e}_{12} \hspace{-1mm} \parallel}^{2} \leq {\epsilon}^{2}_{12} \ \Longrightarrow \ \nonumber \\  
-(\boldsymbol{h}^{0}_{12} + \boldsymbol{e}_{12}) \boldsymbol{\Phi}_{2} (\boldsymbol{h}^{0}_{12} + \boldsymbol{e}_{12})^{\ast} + t_{8} \ \leq \ 0 ,  \label{eqn119} \\
\forall \boldsymbol{e}_{12} \quad \text{s.t.} \quad {\parallel \hspace{-1mm} \boldsymbol{e}_{12} \hspace{-1mm} \parallel}^{2} \leq {\epsilon}^{2}_{12} \ \Longrightarrow \ \nonumber \\ 
(\boldsymbol{h}^{0}_{12} + \boldsymbol{e}_{12}) \boldsymbol{\Psi}_{2} (\boldsymbol{h}^{0}_{12} + \boldsymbol{e}_{12})^{\ast} - t_{10} \ \leq \ 0, \label{eqn120} \\
\forall \boldsymbol{e}_{11} \quad \text{s.t.} \quad {\parallel \hspace{-1mm} \boldsymbol{e}_{11} \hspace{-1mm} \parallel}^{2} \leq {\epsilon}^{2}_{11} \ \Longrightarrow \ \nonumber \\ 
\boldsymbol{e}_{11} (\boldsymbol{\Phi}_{1} + \boldsymbol{\Psi}_{1}) \boldsymbol{e}^{\ast}_{11} - t_{9} \ \leq \ 0, \label{eqn127} \\
\boldsymbol{\Phi}_{1} \succeq \boldsymbol{0}, \quad \boldsymbol{\Psi}_{1} \succeq \boldsymbol{0}, \quad \Tr(\boldsymbol{\Phi}_{1} + \boldsymbol{\Psi}_{1}) \ \leq \ P_{1}, \nonumber \\
\boldsymbol{\Phi}_{2} \succeq \boldsymbol{0}, \quad \boldsymbol{\Psi}_{2} \succeq \boldsymbol{0}, \quad \Tr(\boldsymbol{\Phi}_{2} + \boldsymbol{\Psi}_{2}) \ \leq \ P_{2}.
\label{eqn105}
\end{eqnarray}

\vspace{-2mm}
\hspace{-5mm}
We use the S-procedure to transform the pairs of quadratic inequalities 
in (\ref{eqn111}), (\ref{eqn112}), (\ref{eqn113}), (\ref{eqn114}), 
(\ref{eqn116}), (\ref{eqn117}), (\ref{eqn126}), (\ref{eqn119}), 
(\ref{eqn120}), and (\ref{eqn127}) to equivalent linear matrix 
inequalities (LMI) \cite{ic16}. With this, we get the following 
single minimization form for the above optimization problem:

\vspace{-5mm}
{\small
\begin{eqnarray}
\min_{\boldsymbol{\Phi}_{1}, \boldsymbol{\Psi}_{1}, \boldsymbol{\Phi}_{2}, \boldsymbol{\Psi}_{2}, \atop{t_{1}, t_{2},\cdots,t_{10}, \atop{\lambda_{1}, \lambda_{2},\cdots,\lambda_{10}, \atop{t}}}} \hspace{-1mm} t \hspace{-0mm} \label{eqn106} & 
\\
\text{s.t.} \quad t_{3} \ \geq \ 0, \ t_{4} \ \geq \ 0, \ t_{5} \ \geq \ 0, \ t_{8} \ \geq \ 0, & \nonumber \\
\big( t_{1} + t_{2} \big) - t \big( N_{0} + t_{3} + t_{4}\big) \ \leq \ 0, & \nonumber \\
(2^{R^{'k}_{1}} -1) \big( N_{0} + t_{6} + t_{7} \big) -  t_{5} \ \leq \ 0, & \nonumber \\
(2^{R^{'l}_{2}} -1) \big( N_{0} + t_{9} + t_{10} \big) -  t_{8}  \ \leq \ 0, & \nonumber \\
\left[\footnotesize
\begin{array}{cc}
-\boldsymbol{\Phi}_{1} + \lambda_{1}\boldsymbol{I} & -\boldsymbol{\Phi}_{1}\boldsymbol{z}^{0\ast}_{1} \\ -\boldsymbol{z}^{0}_{1}\boldsymbol{\Phi}^{\ast}_{1} & -\boldsymbol{z}^{0}_{1}\boldsymbol{\Phi}_{1}\boldsymbol{z}^{0\ast}_{1} + t_{1} - \lambda_{1}\epsilon^{2}_{1}
\end{array} \right] \succeq \boldsymbol{0}, \quad \lambda_{1} \geq 0, & \nonumber \\ 
\left[\footnotesize
\begin{array}{cc}
\boldsymbol{\Psi}_{1} + \lambda_{2}\boldsymbol{I} & \boldsymbol{\Psi}_{1}\boldsymbol{z}^{0\ast}_{1} \\ \boldsymbol{z}^{0}_{1}\boldsymbol{\Psi}^{\ast}_{1} & \boldsymbol{z}^{0}_{1}\boldsymbol{\Psi}_{1}\boldsymbol{z}^{0\ast}_{1} - t_{3} - \lambda_{2}\epsilon^{2}_{1}
\end{array} \right] \succeq \boldsymbol{0}, \quad \lambda_{2} \geq 0, & \nonumber \\ 
\left[\footnotesize
\begin{array}{cc}
-\boldsymbol{\Phi}_{2} + \lambda_{3}\boldsymbol{I} & -\boldsymbol{\Phi}_{2}\boldsymbol{z}^{0\ast}_{2} \\ -\boldsymbol{z}^{0}_{2}\boldsymbol{\Phi}^{\ast}_{2} & -\boldsymbol{z}^{0}_{2}\boldsymbol{\Phi}^{}_{2}\boldsymbol{z}^{0\ast}_{2} + t_{2} - \lambda_{3}\epsilon^{2}_{2}
\end{array} \right] \succeq \boldsymbol{0}, \quad \lambda_{3} \geq 0, & \nonumber \\ 
\left[\footnotesize
\begin{array}{cc}
\boldsymbol{\Psi}_{2} + \lambda_{4}\boldsymbol{I} & \boldsymbol{\Psi}_{2}\boldsymbol{z}^{0\ast}_{2} \\ \boldsymbol{z}^{0}_{2}\boldsymbol{\Psi}^{\ast}_{2} & \boldsymbol{z}^{0}_{2}\boldsymbol{\Psi}^{}_{2}\boldsymbol{z}^{0\ast}_{2} - t_{4} - \lambda_{4}\epsilon^{2}_{2}
\end{array} \right] \succeq \boldsymbol{0}, \quad \lambda_{4} \geq 0,  & \nonumber \\ 
\left[\footnotesize
\begin{array}{cc}
\boldsymbol{\Phi}_{1} + \lambda_{5}\boldsymbol{I} & \boldsymbol{\Phi}_{1}\boldsymbol{h}^{0\ast}_{21} \\ \boldsymbol{h}^{0}_{21}\boldsymbol{\Phi}^{\ast}_{1} & \boldsymbol{h}^{0}_{21}\boldsymbol{\Phi}^{}_{1}\boldsymbol{h}^{0\ast}_{21} - t_{5} - \lambda_{5}\epsilon^{2}_{21}
\end{array} \right] \succeq \boldsymbol{0}, \quad \lambda_{5} \geq 0,  & \nonumber \\ 
\left[\footnotesize
\begin{array}{cc}
-\boldsymbol{\Psi}_{1} + \lambda_{6}\boldsymbol{I} & -\boldsymbol{\Psi}_{1}\boldsymbol{h}^{0\ast}_{21} \\ -\boldsymbol{h}^{0}_{21}\boldsymbol{\Psi}^{\ast}_{1} & -\boldsymbol{h}^{0}_{21}\boldsymbol{\Psi}^{}_{1}\boldsymbol{h}^{0\ast}_{21} + t_{7} - \lambda_{6}\epsilon^{2}_{21}
\end{array} \right] \succeq \boldsymbol{0}, \quad \lambda_{6} \geq 0, & \nonumber \\ 
\left[\footnotesize
\begin{array}{cc}
- (\boldsymbol{\Phi}_{2} + \boldsymbol{\Psi}_{2}) + \lambda_{7}\boldsymbol{I} & \boldsymbol{0} \\ \boldsymbol{0} & t_{6} - \lambda_{7}\epsilon^{2}_{22}
\end{array} \right] \succeq \boldsymbol{0}, \quad \lambda_{7} \geq 0, & \nonumber \\ 
\left[\footnotesize
\begin{array}{cc}
\boldsymbol{\Phi}_{2} + \lambda_{8}\boldsymbol{I} & \boldsymbol{\Phi}_{2}\boldsymbol{h}^{0\ast}_{12} \\ \boldsymbol{h}^{0}_{12}\boldsymbol{\Phi}^{\ast}_{2} & \boldsymbol{h}^{0}_{12}\boldsymbol{\Phi}^{}_{2}\boldsymbol{h}^{0\ast}_{12} - t_{8} - \lambda_{8}\epsilon^{2}_{12}
\end{array} \right] \succeq \boldsymbol{0}, \quad \lambda_{8} \geq 0,  & \nonumber \\ 
& \hspace{-87mm} 
\left[\footnotesize
\begin{array}{cc}
-\boldsymbol{\Psi}_{2} + \lambda_{9}\boldsymbol{I} & -\boldsymbol{\Psi}_{2}\boldsymbol{h}^{0\ast}_{12} \\ -\boldsymbol{h}^{0}_{12}\boldsymbol{\Psi}^{\ast}_{2} & -\boldsymbol{h}^{0}_{12}\boldsymbol{\Psi}^{}_{2}\boldsymbol{h}^{0\ast}_{12} + t_{10} - \lambda_{9}\epsilon^{2}_{12}
\end{array} \right] \succeq \boldsymbol{0}, \quad \lambda_{9} \geq 0, \nonumber \\ 
& \hspace{-80mm} 
\left[\footnotesize
\begin{array}{cc}
- (\boldsymbol{\Phi}_{1} + \boldsymbol{\Psi}_{1}) + \lambda_{10}\boldsymbol{I} & \boldsymbol{0} \\ \boldsymbol{0} & t_{9} - \lambda_{10}\epsilon^{2}_{11}
\end{array} \right] \succeq \boldsymbol{0}, \quad \lambda_{10} \geq 0, \nonumber \\ 
& \hspace{-90mm} 
\boldsymbol{\Phi}_{1} \succeq \boldsymbol{0}, \quad \boldsymbol{\Psi}_{1} \succeq \boldsymbol{0}, \quad \Tr(\boldsymbol{\Phi}_{1} + \boldsymbol{\Psi}_{1}) \ \leq \ P_{1}, \nonumber \\
& \hspace{-90mm} \boldsymbol{\Phi}_{2} \succeq \boldsymbol{0}, \quad \boldsymbol{\Psi}_{2} \succeq \boldsymbol{0}, \quad \Tr(\boldsymbol{\Phi}_{2} + \boldsymbol{\Psi}_{2}) \leq \ P_{2}. \hspace{-1mm}
\label{eqn107}
\end{eqnarray}
}

\vspace{-6mm}
\hspace{-4.5mm}
For a given $t$, the above problem is formulated as the following 
semidefinite feasibility problem \cite{ic16}: 
\begin{eqnarray}
\text{find} \quad \boldsymbol{\Phi}_{1}, \ \boldsymbol{\Psi}_{1}, \ \boldsymbol{\Phi}_{2}, \ \boldsymbol{\Psi}_{2}, \ t_{1}, \cdots,t_{10}, %\nonumber \\ 
\ \lambda_{1}, \cdots,\lambda_{10},  \label{eqn108}
\end{eqnarray}
subject to the constraints in (\ref{eqn107}). 
The minimum value of $t$, denoted by $t^{kl}_{min}$, can be obtained using 
bisection method \cite{ic16} as described in Section \ref{sec3}. The value of $t_{lowerlimit}$ 
can be taken as 0 (corresponding to the minimum information rate of 0). 
The value of $t_{upperlimit}$ can be taken as $(2^{C^{'}_{E}} - 1)$, which 
corresponds to the best case information capacity of the eavesdropper link.
Using $t^{kl}_{min}$ in (\ref{eqn92}), the upper bound on $R^{''kl}_{E}$ is given by
\begin{eqnarray}
R^{''kl}_{E} \ \leq \ \log_2 \big(1 + t^{kl}_{min} \big). \label{eqn109}
\end{eqnarray}
Similarly, denoting the optimal values of $t_{5},\cdots,t_{10}$
by $t^{kl}_{5},\cdots,t^{kl}_{10}$, we obtain lower bounds
on $R^{''k}_{1}$ and $R^{''l}_{2}$ as 
\begin{eqnarray}
R^{''k}_{1} \ \geq \ \log_2 \Big(1 + \frac{t^{kl}_{5}}{N_{0} + t^{kl}_{6} + t^{kl}_{7}} \Big), \label{eqn129} \\
R^{''l}_{2} \ \geq \ \log_2 \Big(1 + \frac{t^{kl}_{8}}{N_{0} + t^{kl}_{9} + t^{kl}_{10}} \Big). \label{eqn130}
\end{eqnarray}
Using the upper bound from (\ref{eqn109}) and lower bounds from (\ref{eqn129}) and (\ref{eqn130}),
the lower bound on the worst case sum secrecy rate is 
given by $\max_{k = 0,1,2,\cdots,K, \atop{l = 0,1,2,\cdots,L}} \ (R^{''k}_{1} + R^{''l}_{2} - R^{''kl}_{E})$.

{\em Remark:}
We note that when $S_1$ and $S_2$ do not transmit jamming signals, the 
optimization problems (\ref{eqn102}) and (\ref{eqn104}) will be equivalent, 
and the sum
secrecy rate will be exact. However, the lower bound on the sum secrecy
rate as obtained above with jamming strategies will always be greater
than or equal to the (exact) sum secrecy rate with no jamming 
strategies.

\subsection{Transmit Power Minimization with SINR Constraints} 
\label{sec5}
In this subsection, we minimize the total transmit power (i.e., $S_1$ 
transmit power plus $S_2$ transmit power) with imperfect CSI subject 
to receive SINR constraints at $S_1$, $S_2$, $E$, and individual transmit 
power constraints. The optimization problem to minimize the total transmit 
power is as follows:

\vspace{-5mm}
{\footnotesize
\begin{eqnarray}
\min_{\boldsymbol{\Phi}_{1}, \ \boldsymbol{\Psi}_{1}, \ \boldsymbol{\Phi}_{2}, \ \boldsymbol{\Psi}_{2}} \quad \Tr(\boldsymbol{\Phi}_{1} + \boldsymbol{\Psi}_{1}) +  \Tr(\boldsymbol{\Phi}_{2} + \boldsymbol{\Psi}_{2}) \label{eqn131}  \\
\vspace{2mm}
\text{s.t.} \quad \quad \max_{\boldsymbol{e}_{1}, \ \boldsymbol{e}_{2}} \nonumber \\
\bigg(\frac{ (\boldsymbol{z}^{0}_{1} + \boldsymbol{e}_{1}) \boldsymbol{\Phi}_{1}  (\boldsymbol{z}^{0}_{1} + \boldsymbol{e}_{1} )^{\ast} + (\boldsymbol{z}^{0}_{2} + \boldsymbol{e}_{2} ) \boldsymbol{\Phi}_{2} (\boldsymbol{z}^{0}_{2} + \boldsymbol{e}_{2} )^{\ast} }{ N_{0} + (\boldsymbol{z}^{0}_{1} + \boldsymbol{e}_{1}) \boldsymbol{\Psi}_{1} (\boldsymbol{z}^{0}_{1} + \boldsymbol{e}_{1})^{\ast} + (\boldsymbol{z}^{0}_{2} + \boldsymbol{e}_{2}) \boldsymbol{\Psi}_{2} (\boldsymbol{z}^{0}_{2} + \boldsymbol{e}_{2})^{\ast} } \bigg) \nonumber \\
\leq \ \gamma_{E}, \label{eqn132} \\
\min_{\boldsymbol{e}_{21}, \ \boldsymbol{e}_{22}} \nonumber \\ 
\bigg(\frac{ (\boldsymbol{h}^{0}_{21} + \boldsymbol{e}_{21}) \boldsymbol{\Phi}_{1}  (\boldsymbol{h}^{0}_{21} + \boldsymbol{e}_{21})^{\ast} }{N_{0} + \boldsymbol{e}_{22}(\boldsymbol{\Phi}_{2} + \boldsymbol{\Psi}_{2}) \boldsymbol{e}^{\ast}_{22} + (\boldsymbol{h}^{0}_{21} + \boldsymbol{e}_{21}) \boldsymbol{\Psi}_{1}  (\boldsymbol{h}^{0}_{21} + \boldsymbol{e}_{21})^{\ast} }\bigg) \nonumber \\ 
\geq \ \gamma_{S_{2}},  \label{eqn133} \\
\min_{\boldsymbol{e}_{11}, \ \boldsymbol{e}_{12}} \nonumber \\ 
\bigg(\frac{ (\boldsymbol{h}^{0}_{12} + \boldsymbol{e}_{12}) \boldsymbol{\Phi}_{2}  (\boldsymbol{h}^{0}_{12} + \boldsymbol{e}_{12})^{\ast} }{N_{0} + \boldsymbol{e}_{11}(\boldsymbol{\Phi}_{1} + \boldsymbol{\Psi}_{1}) \boldsymbol{e}^{\ast}_{11} + (\boldsymbol{h}^{0}_{12} + \boldsymbol{e}_{12}) \boldsymbol{\Psi}_{2}  (\boldsymbol{h}^{0}_{12} + \boldsymbol{e}_{12})^{\ast} }\bigg) \nonumber \\ 
\geq \ \gamma_{S_{1}},  \label{eqn134} \\
{\parallel \hspace{-1mm} \boldsymbol{e}_{11} \hspace{-1mm} \parallel}^{2} \leq {\epsilon}^{2}_{11}, \quad 
{\parallel \hspace{-1mm} \boldsymbol{e}_{12} \hspace{-1mm} \parallel}^{2} \leq {\epsilon}^{2}_{12}, \quad
{\parallel \hspace{-1mm} \boldsymbol{e}_{21} \hspace{-1mm} \parallel}^{2} \leq {\epsilon}^{2}_{21}, \nonumber \\
{\parallel \hspace{-1mm} \boldsymbol{e}_{22} \hspace{-1mm} \parallel}^{2} \leq {\epsilon}^{2}_{22}, \quad 
{\parallel \hspace{-1mm} \boldsymbol{e}_{1} \hspace{-1mm} \parallel}^{2} \leq {\epsilon}^{2}_{1}, \quad 
{\parallel \hspace{-1mm} \boldsymbol{e}_{2} \hspace{-1mm} \parallel}^{2} \leq {\epsilon}^{2}_{2},  \nonumber \\
\boldsymbol{\Phi}_{1} \succeq \boldsymbol{0}, \quad \boldsymbol{\Psi}_{1} \succeq \boldsymbol{0}, \quad \Tr(\boldsymbol{\Phi}_{1} + \boldsymbol{\Psi}_{1}) \ \leq \ P_{1}, \nonumber \\
\boldsymbol{\Phi}_{2} \succeq \boldsymbol{0}, \quad \boldsymbol{\Psi}_{2} \succeq \boldsymbol{0}, \quad \Tr(\boldsymbol{\Phi}_{2} + \boldsymbol{\Psi}_{2}) \ \leq \ P_{2}.
\label{eqn135}
\end{eqnarray}
}

\vspace{-5mm}
\hspace{-5mm}
The left hand side of the inequality in the constraint (\ref{eqn132}) 
corresponds to the best case received SINR at the eavesdropper over
the region of CSI error uncertainty. Similarly, the left hand side of 
the inequality in the constraints (\ref{eqn133}) and (\ref{eqn134}) 
correspond to the worst case received SINR at $S_{2}$, and $S_{1}$, 
respectively. $\gamma_{E}$, $\gamma_{S_{2}}$, and $\gamma_{S_{1}}$
are known SINR thresholds at $E$, $S_{2}$, and $S_{1}$, respectively.
Solving the above optimization problem is hard due to the 
presence of $\boldsymbol{e}_{1}$ and $\boldsymbol{e}_{2}$ in both the 
numerator and denominator of the SINR expression of the eavesdropper 
in (\ref{eqn132}). Similarly, $\boldsymbol{e}_{21}$ and 
$\boldsymbol{e}_{12}$ appear in both the numerator and denominator
of the SINR expressions of $S_{2}$ and $S_{1}$ in the constraints 
(\ref{eqn133}) and (\ref{eqn134}), respectively.
By independently constraining the various quadratic terms appearing 
in the constraints (\ref{eqn132}), (\ref{eqn133}), and (\ref{eqn134}), 
and further using the S-procedure, we get the following upper bound 
for the above optimization problem:

\vspace{-5mm}
{\small
\begin{eqnarray}
\min_{\boldsymbol{\Phi}_{1}, \boldsymbol{\Psi}_{1}, \boldsymbol{\Phi}_{2}, \boldsymbol{\Psi}_{2}, \atop{t_{1}, t_{2},\cdots,t_{10}, \atop{\lambda_{1}, \lambda_{2},\cdots,\lambda_{10}}}} \hspace{-1mm} \Tr(\boldsymbol{\Phi}_{1} + \boldsymbol{\Psi}_{1}) +  \Tr(\boldsymbol{\Phi}_{2} + \boldsymbol{\Psi}_{2}) \hspace{-0mm} \label{eqn136} & 
\\
\text{s.t.} \quad t_{3} \ \geq \ 0, \ t_{4} \ \geq \ 0, \ t_{5} \ \geq \ 0, \ t_{8} \ \geq \ 0, & \nonumber \\
\big( t_{1} + t_{2} \big) - \gamma_{E} \big( N_{0} + t_{3} + t_{4}\big) \ \leq \ 0, & \nonumber \\
\gamma_{S_{2}} \big( N_{0} + t_{6} + t_{7} \big) -  t_{5} \ \leq \ 0, & \nonumber \\
\gamma_{S_{1}} \big( N_{0} + t_{9} + t_{10} \big) -  t_{8}  \ \leq \ 0, & \nonumber \\
\left[\footnotesize
\begin{array}{cc}
-\boldsymbol{\Phi}_{1} + \lambda_{1}\boldsymbol{I} & -\boldsymbol{\Phi}_{1}\boldsymbol{z}^{0\ast}_{1} \\ -\boldsymbol{z}^{0}_{1}\boldsymbol{\Phi}^{\ast}_{1} & -\boldsymbol{z}^{0}_{1}\boldsymbol{\Phi}_{1}\boldsymbol{z}^{0\ast}_{1} + t_{1} - \lambda_{1}\epsilon^{2}_{1}
\end{array} \right] \succeq \boldsymbol{0}, \quad \lambda_{1} \geq 0, & \nonumber \\ 
\left[\footnotesize
\begin{array}{cc}
\boldsymbol{\Psi}_{1} + \lambda_{2}\boldsymbol{I} & \boldsymbol{\Psi}_{1}\boldsymbol{z}^{0\ast}_{1} \\ \boldsymbol{z}^{0}_{1}\boldsymbol{\Psi}^{\ast}_{1} & \boldsymbol{z}^{0}_{1}\boldsymbol{\Psi}_{1}\boldsymbol{z}^{0\ast}_{1} - t_{3} - \lambda_{2}\epsilon^{2}_{1}
\end{array} \right] \succeq \boldsymbol{0}, \quad \lambda_{2} \geq 0, & \nonumber \\ 
\left[\footnotesize
\begin{array}{cc}
-\boldsymbol{\Phi}_{2} + \lambda_{3}\boldsymbol{I} & -\boldsymbol{\Phi}_{2}\boldsymbol{z}^{0\ast}_{2} \\ -\boldsymbol{z}^{0}_{2}\boldsymbol{\Phi}^{\ast}_{2} & -\boldsymbol{z}^{0}_{2}\boldsymbol{\Phi}^{}_{2}\boldsymbol{z}^{0\ast}_{2} + t_{2} - \lambda_{3}\epsilon^{2}_{2}
\end{array} \right] \succeq \boldsymbol{0}, \quad \lambda_{3} \geq 0, & \nonumber \\ 
\left[\footnotesize
\begin{array}{cc}
\boldsymbol{\Psi}_{2} + \lambda_{4}\boldsymbol{I} & \boldsymbol{\Psi}_{2}\boldsymbol{z}^{0\ast}_{2} \\ \boldsymbol{z}^{0}_{2}\boldsymbol{\Psi}^{\ast}_{2} & \boldsymbol{z}^{0}_{2}\boldsymbol{\Psi}^{}_{2}\boldsymbol{z}^{0\ast}_{2} - t_{4} - \lambda_{4}\epsilon^{2}_{2}
\end{array} \right] \succeq \boldsymbol{0}, \quad \lambda_{4} \geq 0,  & \nonumber \\ 
\left[\footnotesize
\begin{array}{cc}
\boldsymbol{\Phi}_{1} + \lambda_{5}\boldsymbol{I} & \boldsymbol{\Phi}_{1}\boldsymbol{h}^{0\ast}_{21} \\ \boldsymbol{h}^{0}_{21}\boldsymbol{\Phi}^{\ast}_{1} & \boldsymbol{h}^{0}_{21}\boldsymbol{\Phi}^{}_{1}\boldsymbol{h}^{0\ast}_{21} - t_{5} - \lambda_{5}\epsilon^{2}_{21}
\end{array} \right] \succeq \boldsymbol{0}, \quad \lambda_{5} \geq 0,  & \nonumber \\ 
\left[\footnotesize
\begin{array}{cc}
-\boldsymbol{\Psi}_{1} + \lambda_{6}\boldsymbol{I} & -\boldsymbol{\Psi}_{1}\boldsymbol{h}^{0\ast}_{21} \\ -\boldsymbol{h}^{0}_{21}\boldsymbol{\Psi}^{\ast}_{1} & -\boldsymbol{h}^{0}_{21}\boldsymbol{\Psi}^{}_{1}\boldsymbol{h}^{0\ast}_{21} + t_{7} - \lambda_{6}\epsilon^{2}_{21}
\end{array} \right] \succeq \boldsymbol{0}, \quad \lambda_{6} \geq 0, & \nonumber \\ 
\left[\footnotesize
\begin{array}{cc}
- (\boldsymbol{\Phi}_{2} + \boldsymbol{\Psi}_{2}) + \lambda_{7}\boldsymbol{I} & \boldsymbol{0} \\ \boldsymbol{0} & t_{6} - \lambda_{7}\epsilon^{2}_{22}
\end{array} \right] \succeq \boldsymbol{0}, \quad \lambda_{7} \geq 0, & \nonumber \\ 
\left[\footnotesize
\begin{array}{cc}
\boldsymbol{\Phi}_{2} + \lambda_{8}\boldsymbol{I} & \boldsymbol{\Phi}_{2}\boldsymbol{h}^{0\ast}_{12} \\ \boldsymbol{h}^{0}_{12}\boldsymbol{\Phi}^{\ast}_{2} & \boldsymbol{h}^{0}_{12}\boldsymbol{\Phi}^{}_{2}\boldsymbol{h}^{0\ast}_{12} - t_{8} - \lambda_{8}\epsilon^{2}_{12}
\end{array} \right] \succeq \boldsymbol{0}, \quad \lambda_{8} \geq 0,  & \nonumber \\ 
& \hspace{-87mm} 
\left[\footnotesize
\begin{array}{cc}
-\boldsymbol{\Psi}_{2} + \lambda_{9}\boldsymbol{I} & -\boldsymbol{\Psi}_{2}\boldsymbol{h}^{0\ast}_{12} \\ -\boldsymbol{h}^{0}_{12}\boldsymbol{\Psi}^{\ast}_{2} & -\boldsymbol{h}^{0}_{12}\boldsymbol{\Psi}^{}_{2}\boldsymbol{h}^{0\ast}_{12} + t_{10} - \lambda_{9}\epsilon^{2}_{12}
\end{array} \right] \succeq \boldsymbol{0}, \quad \lambda_{9} \geq 0, \nonumber \\ 
& \hspace{-80mm} 
\left[\footnotesize
\begin{array}{cc}
- (\boldsymbol{\Phi}_{1} + \boldsymbol{\Psi}_{1}) + \lambda_{10}\boldsymbol{I} & \boldsymbol{0} \\ \boldsymbol{0} & t_{9} - \lambda_{10}\epsilon^{2}_{11}
\end{array} \right] \succeq \boldsymbol{0}, \quad \lambda_{10} \geq 0, \nonumber \\ 
& \hspace{-90mm} 
\boldsymbol{\Phi}_{1} \succeq \boldsymbol{0}, \quad \boldsymbol{\Psi}_{1} \succeq \boldsymbol{0}, \quad \Tr(\boldsymbol{\Phi}_{1} + \boldsymbol{\Psi}_{1}) \ \leq \ P_{1}, \nonumber \\
& \hspace{-90mm} \boldsymbol{\Phi}_{2} \succeq \boldsymbol{0}, \quad \boldsymbol{\Psi}_{2} \succeq \boldsymbol{0}, \quad \Tr(\boldsymbol{\Phi}_{2} + \boldsymbol{\Psi}_{2}) \leq \ P_{2}, \hspace{-1mm}
\label{eqn137}
\end{eqnarray}
}

\vspace{-6mm}
\hspace{-4.5mm}
where $t_{1},t_{2},\cdots,t_{10}$ are as defined in the optimization 
problem (\ref{eqn104}). The above problem can be easily solved using 
semidefinite programming techniques.

\section{Results and Discussions}
\label{sec6}
\begin{figure}[h]
\vspace{-3mm}
\includegraphics[totalheight=8.0cm,width=9.25cm]{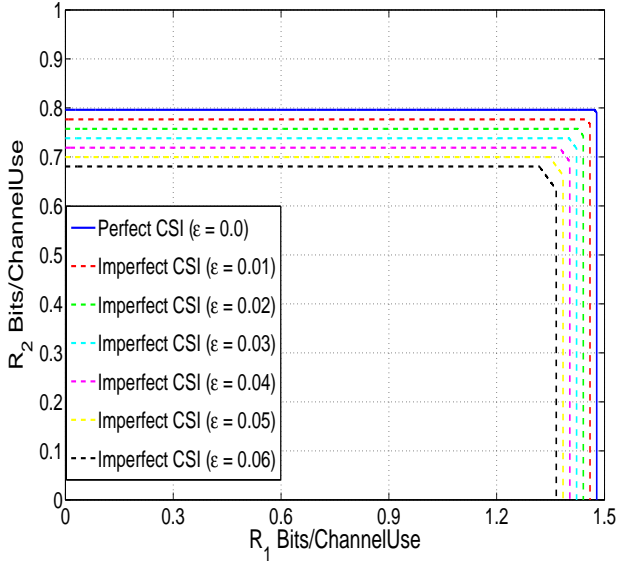}
\vspace{-5mm}
\caption{Achievable $(R_{1},R_{2})$ region in full-duplex communication. 
$P_{1} = P_{2} = 3$ dB, $M_{1} = M_{2} = 2$, $N_0=1$, 
$\epsilon = 0.0, \ 0.01, \ 0.02, \ 0.03, \ 0.04, \ 0.05, \ 0.06$.} 
\label{fig2}
\end{figure}

\begin{figure}[htb]
\includegraphics[totalheight=8.0cm,width=9.25cm]{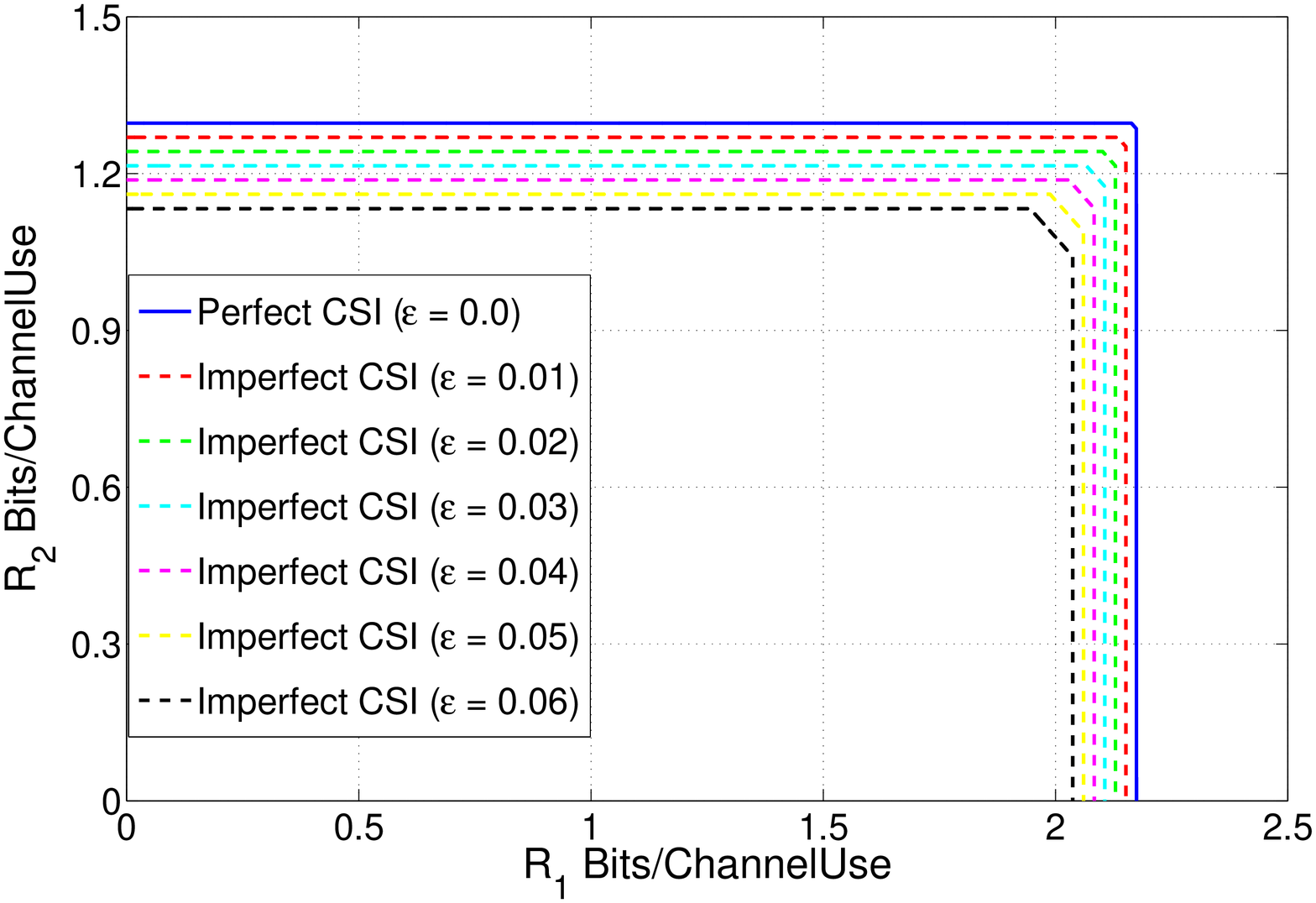}
\vspace{-5mm}
\caption{Achievable $(R_{1},R_{2})$ region in full-duplex communication. 
$P_{1}= P_{2}=6$ dB, $M_{1} = M_{2} = 2$, $N_0=1$,
$\epsilon = 0.0, \ 0.01, \ 0.02, \ 0.03, \ 0.04, \ 0.05, \ 0.06$.} 
\label{fig3}
\end{figure}

In this section, we present numerical results on the secrecy rate
under perfect and imperfect CSI conditions. We assume that 
$M_{1} = M_{2} = 2$. We have used the following channel gains as the 
estimates:
$\boldsymbol{h}^{0}_{12} = [0.0838 + 0.5207i,  \  0.2226 - 0.2482i]$,
$\boldsymbol{h}^{0}_{21} = [0.4407 + 0.6653i,  \  0.5650 - 0.0015i]$,
$\boldsymbol{z}^{0}_{1}  = [0.0765 + 0.0276i,  \  -0.0093 + 0.0062i]$,
$\boldsymbol{z}^{0}_{2}  = [-0.0449 + 0.0314i, \  -0.0396 - 0.0672i]$.
We assume that the magnitudes of the CSI errors in all the links are 
equal, i.e., 
$\epsilon_{11}=\epsilon_{12}=\epsilon_{21}=\epsilon_{22}=\epsilon_{1}=\epsilon_{2}=\epsilon$.
We also assume that $N_{0} = 1$. In Fig. \ref{fig2} and Fig. \ref{fig3},
we plot the $(R_{1},R_{2})$ region obtained by maximizing the sum secrecy 
rate for various values of $\epsilon = 0.0, \ 0.01, \ 0.02, \ 0.03, \ 0.04,
\ 0.05, 0.06$.
Results in Fig. \ref{fig2} and Fig. \ref{fig3} are generated for fixed 
powers $P_{1} = P_{2} = 3$ dB and $P_{1} = P_{2} = 6$ dB, respectively.
We observe that as the magnitude of the CSI error increases the 
corresponding sum secrecy rate decreases which results in the shrinking 
of the achievable rate region. Also, as the power is increased from 
3 dB to 6 dB, the achievable secrecy rate region increases.

\section{Conclusions}
\label{sec7}
We investigated the sum secrecy rate and the corresponding achievable 
secrecy rate region in MISO full-duplex wiretap channel when the CSI in all 
the links were assumed to be imperfect. We obtained the transmit covariance 
matrices associated with the message signals and the jamming signals which 
maximized the worst case sum secrecy rate. Numerical results illustrated 
the impact of imperfect CSI on the achievable secrecy rate region. 
We further note that transmit power optimization subject to outage 
constraint in a slow fading full-duplex MISO wiretap channel can be 
carried out using the approximations by conic optimization in \cite{out1} 
as future extension to this work.

\section*{Appendix A}
In this appendix, we analyze the ranks of the solutions
$\boldsymbol{\Phi}_{1}$, $\boldsymbol{\Psi}_{1}$, $\boldsymbol{\Phi}_{2}$, 
and $\boldsymbol{\Psi}_{2}$ which are obtained by solving the 
optimization problem (\ref{eqn79}) subject to the constraints in 
(\ref{eqn80}). We take the Lagrangian of the objective function $t$ 
subject to the constraints in (\ref{eqn80}) as follows \cite{ic16}:

\vspace{-4mm}
\hspace{-5mm}
{\small
\begin{eqnarray}
\ell(t,\boldsymbol{\Phi}_{1},\boldsymbol{\Psi}_{1},\boldsymbol{\Phi}_{2},\boldsymbol{\Psi}_{2},\lambda_{1},\lambda_{2},\boldsymbol{A}_{1},\boldsymbol{B}_{1},\boldsymbol{A}_{2},\boldsymbol{B}_{2},\mu_{},\nu_{1}, \nu_{2})  = & & \nonumber \\
& \hspace{-70mm} & \hspace{-80mm} \ t + \lambda_{1} \big(  \Tr(\boldsymbol{\Phi}_{1} + \boldsymbol{\Psi}_{1}) - P_{1}  \big)    %\nonumber \\
+ \ \lambda_{2} \big(  \Tr(\boldsymbol{\Phi}_{2} + \boldsymbol{\Psi}_{2}) - P_{2}  \big)             \nonumber \\
&\hspace{-70mm} & \hspace{-80mm} - \ \Tr(\boldsymbol{A}_{1} \boldsymbol{\Phi}_{1}) - \Tr(\boldsymbol{B}_{1} \boldsymbol{\Psi}_{1})  %\nonumber \\
- \ \Tr(\boldsymbol{A}_{2} \boldsymbol{\Phi}_{2}) - \Tr(\boldsymbol{B}_{2} \boldsymbol{\Psi}_{2})  \nonumber \\
&\hspace{-70mm} &\hspace{-80mm} + \ \mu \big( (\boldsymbol{z}_{1} \boldsymbol{\Phi}_{1}  \boldsymbol{z}^{\ast}_{1} + \boldsymbol{z}_{2} \boldsymbol{\Phi}_{2} \boldsymbol{z}^{\ast}_{2} ) - t ( N_{0} + \boldsymbol{z}_{1} \boldsymbol{\Psi}_{1}  \boldsymbol{z}^{\ast}_{1} + \boldsymbol{z}_{2} \boldsymbol{\Psi}_{2} \boldsymbol{z}^{\ast}_{2} )\big )  \nonumber \\
&\hspace{-70mm} &\hspace{-80mm}  + \ \nu_{1} \big( \big( 2^{R^{'k}_{1}} - 1 \big) \big( N_{0} + \boldsymbol{h}_{21} \boldsymbol{\Psi}_{1}  \boldsymbol{h}^{\ast}_{21} \big) - \big( \boldsymbol{h}_{21} \boldsymbol{\Phi}_{1}  \boldsymbol{h}^{\ast}_{21} \big) \big)  \nonumber \\
&\hspace{-70mm} & \hspace{-80mm} + \ \nu_{2} \big( \big( 2^{R^{'l}_{2}} \hspace{-1.0mm} - \hspace{-0.5mm} 1 \big) \big( N_{0} \hspace{-0.5mm} + \boldsymbol{h}_{12} \boldsymbol{\Psi}_{2}  \boldsymbol{h}^{\ast}_{12} \big) \hspace{-0.5mm} - \hspace{-0.5mm} \big( \boldsymbol{h}_{12} \boldsymbol{\Phi}_{2}  \boldsymbol{h}^{\ast}_{12} \big) \big ), \hspace{-1mm}
\label{eqn122}
\end{eqnarray}
}

\vspace{-4mm}
\hspace{-4.5mm}
where $\lambda_{1} \ge 0$, $\lambda_{2} \ge 0$, 
$\boldsymbol{A}_{1} \succeq \boldsymbol{0}$,
$\boldsymbol{B}_{1} \succeq \boldsymbol{0}$, 
$\boldsymbol{A}_{2} \succeq \boldsymbol{0}$,
$\boldsymbol{B}_{2} \succeq \boldsymbol{0}$, 
$\mu \ge 0$, $\nu_{1} \ge 0$, and $\nu_{2} \ge 0$
are the Lagrangian multipliers. The KKT conditions of 
(\ref{eqn122}) are as follows:

\begin{itemize}
\item[] 
\hspace{-8mm}
(a1) all the constraints in (\ref{eqn80}), 
\vspace{1mm}
\item[] 
\hspace{-8mm}
(a2)
$\lambda_{1} \left(  \Tr(\boldsymbol{\Phi}_{1} + \boldsymbol{\Psi}_{1}) - P_{1}  \right) \ = \ 0$,
\vspace{1mm}
\item[] 
\hspace{-8mm}
(a3)
$\lambda_{2} \left(  \Tr(\boldsymbol{\Phi}_{2} + \boldsymbol{\Psi}_{2}) - P_{2}  \right) \ = \ 0$,
\vspace{1mm}
\item[] 
\hspace{-8mm}
(a4)
$\Tr(\boldsymbol{A}_{1} \boldsymbol{\Phi}_{1}) \ = \ 0$. 
Since $\boldsymbol{A}_{1} \ \succeq \ \boldsymbol{0}$ and $\boldsymbol{\Phi}_{1} \ \succeq \ \boldsymbol{0}$ $ \ \Longrightarrow$ $\boldsymbol{A}_{1} \boldsymbol{\Phi}_{1} \ = \ \boldsymbol{0}$,
\item[] 
\hspace{-8mm}
(a5)
$\Tr(\boldsymbol{B}_{1} \boldsymbol{\Psi}_{1}) \ = \ 0$. 
Since $\boldsymbol{B}_{1} \ \succeq \ \boldsymbol{0}$ and $\boldsymbol{\Psi}_{1} \ \succeq \ \boldsymbol{0}$ $ \ \Longrightarrow$ $\boldsymbol{B}_{1} \boldsymbol{\Psi}_{1} \ = \ \boldsymbol{0}$,
\vspace{1mm}
\item[] 
\hspace{-8mm}
(a6)
$\Tr(\boldsymbol{A}_{2} \boldsymbol{\Phi}_{2}) \ = \ 0$. 
Since $\boldsymbol{A}_{2} \ \succeq \ \boldsymbol{0}$ and $\boldsymbol{\Phi}_{2} \ \succeq \ \boldsymbol{0}$ $ \ \Longrightarrow$ $\boldsymbol{A}_{2} \boldsymbol{\Phi}_{2} \ = \ \boldsymbol{0}$,
\vspace{1mm}
\item[] 
\hspace{-8mm}
(a7)
$\Tr(\boldsymbol{B}_{2} \boldsymbol{\Psi}_{2}) \ = \ 0$. 
Since $\boldsymbol{B}_{2} \ \succeq \ \boldsymbol{0}$ and $\boldsymbol{\Psi}_{2} \ \succeq \ \boldsymbol{0}$ $ \ \Longrightarrow$ $\boldsymbol{B}_{2} \boldsymbol{\Psi}_{2} \ = \ \boldsymbol{0}$,
\vspace{1mm}
\item[]
\hspace{-8mm}
(a8)
{\small
$\mu \big( (\boldsymbol{z}_{1} \boldsymbol{\Phi}_{1}  \boldsymbol{z}^{\ast}_{1} + \boldsymbol{z}_{2} \boldsymbol{\Phi}_{2} \boldsymbol{z}^{\ast}_{2} ) - t ( N_{0} + \boldsymbol{z}_{1} \boldsymbol{\Psi}_{1}  \boldsymbol{z}^{\ast}_{1} + \boldsymbol{z}_{2} \boldsymbol{\Psi}_{2} \boldsymbol{z}^{\ast}_{2} )\big ) = 0$,
}
\vspace{-3mm}
\item[]
\hspace{-8mm}
(a9)
$\nu_{1} \big( \big( 2^{R^{'k}_{1}} - 1 \big) \big( N_{0} + \boldsymbol{h}_{21} \boldsymbol{\Psi}_{1}  \boldsymbol{h}^{\ast}_{21} \big) - \big( \boldsymbol{h}_{21} \boldsymbol{\Phi}_{1}  \boldsymbol{h}^{\ast}_{21} \big) \big) = 0$,
\vspace{1mm}
\item[]
\hspace{-8mm}
(a10)
{\small 
$\nu_{2} \big( \big( 2^{R^{'l}_{2}} - 1 \big) \big( N_{0} + \boldsymbol{h}_{12} \boldsymbol{\Psi}_{2}  \boldsymbol{h}^{\ast}_{12} \big) - \big( \boldsymbol{h}_{12} \boldsymbol{\Phi}_{2}  \boldsymbol{h}^{\ast}_{12} \big) \big ) = 0$,
}
\vspace{1mm}
\item[] 
\hspace{-8mm}
(a11) $\frac{\partial \ell}{\partial t} \ = \ 0$ $ \ \Longrightarrow \ $
$\mu \big( N_{0} + \boldsymbol{z}_{1} \boldsymbol{\Psi}_{1}  \boldsymbol{z}^{\ast}_{1} + \boldsymbol{z}_{2} \boldsymbol{\Psi}_{2} \boldsymbol{z}^{\ast}_{2} \big ) = 1$. This implies
that $\mu > 0$,
\vspace{1mm}
\item[] 
\hspace{-8mm}
(a12) $\frac{\partial \ell}{\partial \boldsymbol{\Phi}_{1}} \ = \ \boldsymbol{0}$ $ \ \Longrightarrow \ $
$\boldsymbol{A}_{1} = \lambda_{1} \boldsymbol{I} + \mu \boldsymbol{z}^{\ast}_{1} \boldsymbol{z}_{1} - \nu_{1} \boldsymbol{h}^{\ast}_{21} \boldsymbol{h}_{21} \succeq \boldsymbol{0}$,
\vspace{0mm}
\item[] 
\hspace{-8mm}
(a13) $\frac{\partial \ell}{\partial \boldsymbol{\Psi}_{1}} \ = \ \boldsymbol{0}$ $ \ \Longrightarrow \ $
$\boldsymbol{B}_{1} = \lambda_{1} \boldsymbol{I} - \mu t \boldsymbol{z}^{\ast}_{1} \boldsymbol{z}_{1} + \nu_{1} \big( 2^{R^{'k}_{1}} - 1 \big) \boldsymbol{h}^{\ast}_{21} \boldsymbol{h}_{21} \succeq \boldsymbol{0}$,
\vspace{1mm}
\item[] 
\hspace{-8mm}
(a14) $\frac{\partial \ell}{\partial \boldsymbol{\Phi}_{2}} \ = \ \boldsymbol{0}$ $ \ \Longrightarrow \ $
$\boldsymbol{A}_{2} = \lambda_{2} \boldsymbol{I} + \mu \boldsymbol{z}^{\ast}_{2} \boldsymbol{z}_{2} - \nu_{2} \boldsymbol{h}^{\ast}_{12} \boldsymbol{h}_{12} \succeq \boldsymbol{0}$,
\vspace{0mm}
\item[] 
\hspace{-8mm}
(a15) $\frac{\partial \ell}{\partial \boldsymbol{\Psi}_{2}} \ = \ \boldsymbol{0}$ $ \ \Longrightarrow \ $
$\boldsymbol{B}_{2} = \lambda_{2} \boldsymbol{I} - \mu t \boldsymbol{z}^{\ast}_{2} \boldsymbol{z}_{2} + \nu_{2} \big( 2^{R^{'l}_{2}} - 1 \big) \boldsymbol{h}^{\ast}_{12} \boldsymbol{h}_{12} \succeq \boldsymbol{0}$.
\end{itemize}
We first consider the scenario when $\lambda_{1} > 0$. 
The KKT condition (a12) implies that 
\begin{eqnarray}
\boldsymbol{A}_{1} + \nu_{1} \boldsymbol{h}^{\ast}_{21} \boldsymbol{h}_{21} = \lambda_{1} \boldsymbol{I} + \mu \boldsymbol{z}^{\ast}_{1} \boldsymbol{z}_{1}  \succ \boldsymbol{0}. \label{eqn123}
\end{eqnarray}
The above expression implies that 
$rank(\boldsymbol{A}_{1}) \geq M_{1} - rank\big( \nu_{1} \boldsymbol{h}^{\ast}_{21} \boldsymbol{h}_{21}\big)$.
Since $rank(\nu_{1} \boldsymbol{h}^{\ast}_{21} \boldsymbol{h}_{21}) \leq 1$, 
this further implies that
$rank(\boldsymbol{A}_{1}) \geq M_{1} - 1$. Assuming 
$\boldsymbol{\Phi}_{1} \neq \boldsymbol{0}$, the
KKT condition (a4) implies that $rank(\boldsymbol{A}_{1}) = M_{1}-1$, 
and the expression (\ref{eqn123}) implies that $\nu_{1}>0$. This means 
that $rank(\boldsymbol{\Phi}_{1}) = 1$. With $\lambda_{1} >0$, and 
$\nu_{1} >0$, we rewrite the KKT condition (a13) in the following form:
\begin{eqnarray}
\boldsymbol{B}_{1} + \mu t \boldsymbol{z}^{\ast}_{1} \boldsymbol{z}_{1} = \lambda_{1} \boldsymbol{I}  + \nu_{1} \big( 2^{R^{'k}_{1}} - 1 \big) \boldsymbol{h}^{\ast}_{21} \boldsymbol{h}_{21} \succ \boldsymbol{0}. \label{eqn128} 
\end{eqnarray}
If $t >0$, the above expression implies that 
$rank \big(\boldsymbol{B}_{1} \big) \geq M_{1} - rank\big(\mu t \boldsymbol{z}^{\ast}_{1} \boldsymbol{z}_{1} \big)$ $= M_{1} - 1$.
The KKT condition (a5) implies that 
$rank \big(\boldsymbol{B}_{1} \big) = M_{1} -1$, and 
$rank \big(\boldsymbol{\Psi}_{1} \big) = 1$ 
(assuming $\boldsymbol{\Psi}_{1} \neq \boldsymbol{0}$). Now, if 
$t = 0$, the KKT condition (a8) implies that
$\boldsymbol{z}_{1} \boldsymbol{\Phi}_{1}  \boldsymbol{z}^{\ast}_{1} + \boldsymbol{z}_{2} \boldsymbol{\Phi}_{2} \boldsymbol{z}^{\ast}_{2} = 0$,
i.e., the received signal power at the eavesdropper will be zero. The 
expression (\ref{eqn128}), and the KKT condition (a5) further imply 
that $\boldsymbol{\Psi}_{1} = \boldsymbol{0}$. Also, when $\lambda_{1}>0$,
the KKT condition (a2) implies that 
$\Tr(\boldsymbol{\Phi}_{1} + \boldsymbol{\Psi}_{1}) = P_{1}$, i.e., the 
entire power $P_{1}$ is used for the transmission. Similar rank analysis 
holds for $\boldsymbol{\Phi}_{2}$ and
$\boldsymbol{\Psi}_{2}$ when $\lambda_{2} >0$.

We now consider the scenario when $\lambda_{1} = 0$. 
Assuming $\boldsymbol{z}_{1}$ and $\boldsymbol{h}_{21}$ are not collinear,
the KKT condition (a12) will be satisfied only when $\nu_{1}=0$.
With this, the expression (\ref{eqn123}) implies that 
$\boldsymbol{A}_{1} = \mu \boldsymbol{z}^{\ast}_{1} \boldsymbol{z}_{1}$ 
and 
$rank(\boldsymbol{A}_{1})=rank(\mu \boldsymbol{z}^{\ast}_{1} \boldsymbol{z}_{1})=1$.
The KKT condition (a4) further implies that the eigen vectors 
corresponding to the non-zero eigen values of $\boldsymbol{\Phi}_{1}$
lie in the orthogonal complement subspace of $\boldsymbol{z}^{\ast}_{1}$,
and $rank(\boldsymbol{\Phi}_{1}) \leq M_{1} - 1$.
Further, with $\lambda_{1} = 0$ and $\nu_{1} = 0$, the KKT condition 
$(a13)$ will be satisfied only when $t = 0$ i.e., 
$\boldsymbol{z}_{1} \boldsymbol{\Phi}_{1}  \boldsymbol{z}^{\ast}_{1} + \boldsymbol{z}_{2} \boldsymbol{\Phi}_{2} \boldsymbol{z}^{\ast}_{2} = 0$.
The above analysis implies that there exist a rank-1 optimum 
$\boldsymbol{\Phi}_{1}$.
Similar rank analysis holds for $\boldsymbol{\Phi}_{2}$ and
$\boldsymbol{\Psi}_{2}$ when $\lambda_{2} = 0$.

\end{document}